\documentclass[journal]{IEEEtran}

\ifCLASSINFOpdf
 
\else
  \fi
  \usepackage{subfigure}  \usepackage{adjustbox}

\usepackage[T1]{fontenc}
\usepackage[utf8]{inputenc}
\usepackage{tabularx,ragged2e,booktabs,caption}
\usepackage{mathrsfs}
\newcolumntype{C}[1]{>{\Centering}m{#1}}

\usepackage{graphicx}

\usepackage{amsthm}

\usepackage{tabulary}
\usepackage{amsmath,cite,multirow,amsfonts}
\usepackage{graphicx}
\usepackage{xcolor}
\usepackage{bm}

\usepackage{booktabs}
\usepackage{algorithm}
\usepackage{algcompatible}

\usepackage{algpseudocode}

\usepackage{amsfonts}
\usepackage{microtype}
\usepackage{color}
\usepackage{soul}
\usepackage{ifthen}
\usepackage{textcomp}
\usepackage{mathtools}
\usepackage[hidelinks]{hyperref}

\usepackage{subfigure}

\begin{document}

\title{Evolving Multi-Resolution Pooling CNN  for Monaural Singing Voice Separation}

\author{Weitao Yuan,  Bofei Dong, Shengbei Wang, 
 Masashi Unoki, and Wenwu Wang, ~\IEEEmembership{Senior Member,~IEEE}
\thanks{Weitao Yuan,  Bofei Dong and Shengbei Wang  are with Tianjin Key Laboratory of Autonomous Intelligence Technology and Systems,   School of Computer Science and Technology, Tianjin Polytechnic University, Tianjin, China, e-mail: weitaoyuan@hotmail.com, tuku@tuxzz.org, wangshengbei@tjpu.edu.cn.}
\thanks{Masashi Unoki is with the School of  Information Science, Japan Advanced Institute of Science and Technology, Japan, e-mail: unoki@jaist.ac.jp.}
\thanks{Wenwu Wang is with the Centre for Vision, Speech and Signal Processing, University of Surrey, Guildford, UK, e-mail: w.wang@surrey.ac.uk.}
\vspace{-4mm}
}
%
\maketitle
\begin{abstract} 

Monaural Singing Voice Separation (MSVS) is a challenging task and has been studied for decades. Deep neural networks (DNNs) are the current state-of-the-art methods for MSVS. However, the existing DNNs are often designed manually, which is time-consuming and error-prone. In addition, the network architectures are usually pre-defined, and not adapted to the training data. To address these issues, we introduce a Neural Architecture Search (NAS) method to the structure design of DNNs for MSVS. Specifically, we propose a new multi-resolution Convolutional Neural Network (CNN) framework for MSVS namely Multi-Resolution Pooling CNN (MRP-CNN), which uses various-size pooling operators to extract multi-resolution features. Based on the NAS, we then develop an evolving framework namely Evolving MRP-CNN (E-MRP-CNN), by automatically searching the effective MRP-CNN structures using genetic algorithms, optimized in terms of a single-objective considering only separation performance, or multi-objective considering both the separation performance and the model complexity. The multi-objective E-MRP-CNN gives a set of Pareto-optimal solutions, each providing a trade-off between separation performance and model complexity. Quantitative and qualitative evaluations on the MIR-1K and DSD100 datasets are used to demonstrate the advantages of the proposed  framework over several recent baselines.

\end{abstract}


 \ifCLASSOPTIONpeerreview
 \begin{center} \bfseries EDICS Category: 3-BBND \end{center}
 \fi
\IEEEpeerreviewmaketitle

\vspace{-2mm}
\section{Introduction}\label{sec:intro}

Popular music, which  plays a central role in entertainment industries, usually consists of two components:  singing voice (Vocal) and music accompaniment (Acc)~\cite{RafiiLSMFP18}. Human beings can easily hear out/distinguish the singing voice from music accompaniment when listening to a popular song. This effortless task for human, however,  is very difficult for machines, which raises both challenges and opportunities to advance audio signal processing techniques~\cite{LiW07, RafiiLSMFP18}. Monaural singing voice separation (MSVS), as an important research branch of music source separation (MSS), aims to separate the singing voice and the background music accompaniment from a single-channel mixture signal. The research on MSVS is useful in many areas such as automatic lyrics recognition/alignment, singer identification, and music information retrieval~\cite{LiW07}. Moreover, it  would  benefit our   understanding   of the  perception and interpretation mechanisms of the human auditory system~\cite{LiW07}.

Traditional (largely unsupervised) methods have provided many effective frameworks for MSVS~\cite{RafiiLSMFP18}, e.g., time-frequency (T-F) masking methods~\cite{LiW07}, and robust principal component analysis (RPCA) based methods~\cite{HuangCSH12}. A comprehensive overview of the traditional MSVS methods can be found in~\cite{RafiiLSMFP18}. Benefiting from these methods, recent data-driven methods, especially the Deep Neural Network (DNN)~\cite{DP}, strongly boosts the performance of MSVS with the help of large scale data. The  basic building blocks of DNNs for MSVS mainly include Feed-Forward Network (FFN)~\cite{FFN}, Recurrent Neural Network (RNN)~\cite{HuangKHS15},   Convolutional Neural Network (CNN)~\cite{CNN2},  and attention mechanism~\cite{weitao1}. In these building blocks, CNN is proven to be very effective in extracting vocal/musical features for MSVS, since efficient representations related to discriminative features of vocal/music can be learned by convolutional filters via sharing weights.

 In fact,  music   relies heavily on its multi-scale repetitions (e.g., from very basic elements such as individual notes, timber, or pitch to larger structure  chords~\cite{PaulusMK10}) to build the logical structure and meaning~\cite{musictransformer}. These multi-resolution repetitions appearing at various musical levels also distinguish the music accompaniment from vocals which are less redundant and mostly harmonic~\cite{RafiiLSMFP18}. As an important CNN for  MSVS, the Multi-Resolution CNN (MR-CNN) \cite{SHN, jansson2017singing, DMP1, DMP2}, which can capture multi-resolution features via constructing various-size receptive fields (RFs),  has been found  effective  in modeling the multi-scale repetitive music structures and  extracting discriminative features (e.g., global or local features). The MR-CNN  has been widely employed by many state-of-the-art (SOTA) MSVS methods and it is also our research focus in  this work.

According to different implementations  of multi-resolution RFs, existing MR-CNNs for MSVS/MSS  can be divided into two types. The first type, e.g., Stacked Hourglass Network (SHN)~\cite{SHN}  and U-net~\cite{jansson2017singing}, is constructed in a cascade manner with fixed-size or single-resolution RF in each layer. The input signal is repeatedly convoluted and   downsampled  to  form multiple consecutive layers. In this case, different resolution features can only be found in different layers and  thus the cascade structure  of the first type MR-CNN should be deep enough to extract effective multi-resolution features. In contrast, the second type   MR-CNN such as Multi-Resolution Convolutional Auto-Encoder (MRCAE) \cite{DMP1} and Multi-Resolution Fully Convolutional Neural Network (MR-FCNN) \cite{DMP2}, directly implements multi-resolution RFs in  the same layer by using multiple sets of  various-size  convolutional operators. Accordingly,  multi-resolution features can be extracted in one or a few layers  without  deepening the  cascade structure.  

In spite of these  achievements, several issues need to be addressed for  current MR-CNNs: 
\paragraph*{(1) Architecture limitations}
 
The  first type MR-CNN depends on its cascade structure to extract multi-scale music features. However, according to \cite{SAGAN},  the  optimization algorithms would be  less effective in capturing the dependencies across multiple layers. This  problem could be aggravated in the first type MR-CNN since it heavily relies on its deep cascade structure to improve the separation performance of MSVS. 
 
In contrast, the second type MR-CNN   does not suffer from the optimization issue. However, in order to extract global features, large-size  convolutional filters should be used.  According to~\cite{abs-1902-01492},   large  convolutional filter results in  low  computational efficiency. Moreover, for MSVS, a minor linear shift in T-F representations (e.g., magnitude spectrogram) could cause significant distortions on vocal and music perception~\cite{jansson2017singing}. To  address this issue,  many MSVS networks employ skip or similar connections to directly transmit the low-level information between different layers~\cite{jansson2017singing,SHN}. However,   such skip connections (or similar mechanisms) have not been implemented for the second type MR-CNN.

\paragraph*{(2) Manual  design} 
Current  MR-CNNs (or DNNs)  based MSVS methods are usually designed  manually. This  manual design procedure usually has the following shortcomings. 
 
\begin{enumerate}
 \item Manual design is often achieved empirically via trial and error: MSVS is a challenging task as the music accompaniment and vocals often exhibit highly synchronous non-stationary spectro-temporal structures  over  time and frequency~\cite{RafiiLSMFP18}. The MR-CNN learns hierarchical feature extractors (e.g., the coefficients of the convolutional operators) from the data in an ``end-to-end''  fashion. In this case, slight modifications to the architecture may significantly affect the separation performance. To find suitable structures for MSVS, a large amount of architecture modifications and  repetitive training and testing are required, which  is inevitably time-consuming, error-prone, and ineffective. 
 \item 
Domain knowledge may be not  sufficient for detailed architecture design: For MSVS,  domain knowledge may suggest to use vertical filters to learn timbral representations~\cite{LeePLN09} and  horizontal filters to learn long  temporal cues~\cite{SchluterB14} in the T-F domain.  However,  when dealing with an actual MSVS network, how to combine and deploy these filters and how to select   an effective combination from so many possible combinations may not be answered sufficiently by domain/expert knowledge.
\item  Pre-designed structures lack a mechanism to adapt their architectures to the training data: The data-driven optimization process of MR-CNNs can learn   parameters of the convolutional filters. However,  the pre-defined   convolutional operator sizes,  the  hyper-paremters, and the architecture of MR-CNNs, cannot be changed or adapted to the dataset during the  training process. As a result, the information learned from real data is not utilized for improving the pre-designed structures.
\end{enumerate} 
 
To address these issues, this paper proposes a flexible and effective MR-CNN  for MSVS namely Multi-Resolution Pooling CNN (MRP-CNN). We also extend the proposed MRP-CNN into an evolving framework, i.e., E-MRP-CNN, using Neural Architecture Search (NAS) technique. The E-MRP-CNN can automatically evolve its neural architecture according to  the learned data    using two kinds of   genetic algorithms: the single-objective  genetic algorithm and the multi-objective genetic algorithm. The details of  our work are described below. 
\paragraph*{(1) Multi-resolution Pooling CNN}

The MRP-CNN utilizes sets of average pooling operators of various sizes in parallel at the same layer to obtain multi-resolution features. All these pooling operators are embedded in stacked convolution networks with small and fixed-size convolutional kernels. Compared with the cascade framework U-net or SHN (the first type MR-CNN), the MRP-CNN  does not need to optimize the deep cascade structure. Compared with  the second type MR-CNN, large-size pooling (downsampling) operators rather than  large-size convolutional filters are used to extract global features, which reduces the number of trainable parameters and leads to much   better memory  and  computational efficiency. Moreover, the MRP-CNN  is a flexible design   and  allows skip connections (or other similar connections) to be implemented  between different layers for low-level features transmission.
  
\paragraph*{(2) Automatic Neural Architecture Search}
 We  introduce  NAS  to the MRP-CNN and  construct the E-MRP-CNN,  which can  automatically  search effective MRP-CNN architectures  for MSVS. As the first attempt to  introduce NAS in the MSVS field, we aim to enhance the existing MR-CNNs and make the DNN  based MSVS methods less dependent on domain/expert knowledge, with single-objective E-MRP-CNN and multi-objective E-MRP-CNN.

The single-objective E-MRP-CNN  evolves its architecture with the only objective of optimizing the separation performance. This  evolving process will provide an insight about how  different architectures of MRP-CNN affect the separation performance and what structures work well on the  MSVS problem. The single-objective E-MRP-CNN  tends to optimize the separation performance, but choosing a more complex model. In some real applications (e.g., the embedded FPGA platform)~\cite{GuoZYWY18}, however, the computing resources and  on-chip memory are usually limited, in this case, both the model complexity and separation performance should be considered.
 
The multi-objective E-MRP-CNN is proposed to address the balance between model complexity and  separation performance.  It  provides  a set of Pareto-optimal solutions \cite{DebAPM02} for MSVS, i.e., Pareto-optimal  MRP-CNN architectures. Each solution (architecture) is Pareto-optimal, that is, no objective can be improved without degrading the other objective, e.g.,  the separation performance can not be improved without increasing the  model complexity. We approximate the  Pareto-optimal solution set based on a classic multi-objective evolutionary genetic algorithm: Non-dominated Sorting Genetic Algorithm II (NSGA-II)~\cite{DebAPM02}. With the multi-objective E-MRP-CNN, we can  obtain multiple architectures with each providing a good separation performance under a fixed model complexity.

Our main contributions are summarized as follow. 
\begin{itemize}
\item We propose a flexible MR-CNN framework, i.e., MRP-CNN, for extracting multi-resolution spectro-temporal features for MSVS;  
\item Based on MRP-CNN, we introduce the first evolutionary scheme  for MSVS, i.e.,  the E-MRP-CNN, which can evolve its  architecture and search effective architecture  for MSVS based on training data. This automatic scheme not only avoids the empirical manual design process but  also provides better separation performance  (via the single-objective E-MRP-CNN)  and a well-balanced model complexity and separation performance  (via the multi-objective E-MRP-CNN) for  MSVS.
\end{itemize}


\begin{figure}[t]
\center
\includegraphics[width=0.48\textwidth]{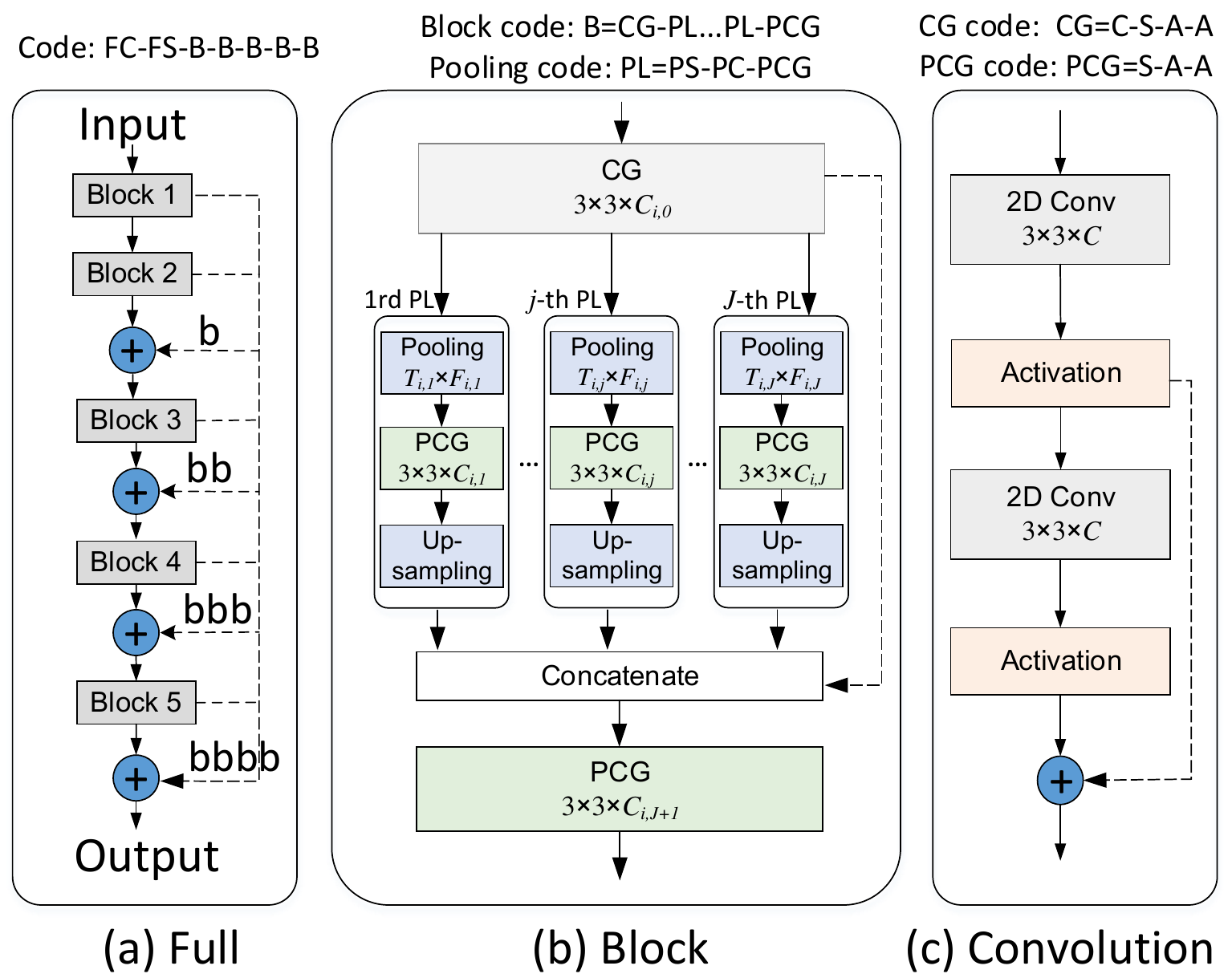} 
\caption{The architecture of the proposed MRP-CNN.}
\label{Block}
\vspace{-2mm}
\end{figure}

\vspace{-2mm}

\section{related works}\label{related}

The existing deep networks for MSVS/MSS mainly use RNN~\cite{MimilakisDSSVB18, MimilakisDVS17} and CNN structures~\cite{CNN2, SHN, jansson2017singing, StollerED18, SlizovskaiaKHG19}. The RNN can effectively model dependencies of temporal patterns and structures of music (e.g. rhythm, beat/tempo, melody)~\cite{MimilakisDSSVB18, MimilakisDVS17}. The CNN, which is effective for feature extraction in the T-F domain, is usually constructed as a convolutional encoder-decoder architecture with skip connections, such as the U-net~\cite{jansson2017singing}, Wave-U-net~\cite{StollerED18}, Exp-Wave-U-Net~\cite{SlizovskaiaKHG19}, and SHN~\cite{SHN}. The CNN can also be combined with other structures to obtain better MSS/MSVS performance. For example, in~\cite{ LiuY19}, CNN and RNN are combined to improve the MSS performance; the Skip Attention (SA)~\cite{weitao1} inspired from Transformer~\cite{transformer} was  introduced into CNN encoder-decoder structure to improve the separation performance. In addition to these works,~\cite{MichelashviliBW19, SubakanS18} considered  using  generative adversarial networks (GANs) for (semi-supervised) MSVS;~\cite{LuoCHRM17} designed Chimera network  for singing voice separation based on deep clustering; \cite{abs-1904-06157} examined the mapping functions of neural networks based on the denoising autoencoder (DAE) model for MSS. However, these works lack flexibility for adapting the architectures to the data, as compared with the use of various size pooling operators in our approach. In addition, all of these networks are designed manually and none of them considered the use of NAS for automatic architecture design.

Over the past few years, the NAS has achieved impressive progress in many research areas and begun to outperform human-designed deep models~\cite{SoLL19, ElskenMH19}. As a classic search strategy of NAS, the NeuroEvolution of Augmenting Topologies (NEAT)~\cite{Stanley2001EvolvingNN} adopted the Genetic Algorithm (GA) to evolve both its artificial neural networks and their weights. Recently, the Evolved Transformer~\cite{SoLL19} considered the use of NAS to find a better alternative to the Transformer for sequence-to-sequence tasks. The Reinforcement Learning (RL) based NAS has also been introduced to Generative Adversarial Networks (GANs)~\cite{abs-1908-03835}. However, to our knowledge, the NAS has not been explored for the MSVS/MSS tasks and no work has yet attempted to design an evolving MR-CNN framework for MSVS/MSS. In particular, since the neural architecture for MSVS usually has millions of weights, we use GA to optimize the neural architecture while the gradient based method to optimize the weights~\cite{ElskenMH19}, which is different from NEAT~\cite{Stanley2001EvolvingNN}. In addition, compared with the RL based NAS (e.g.,~\cite{abs-1908-03835}), the evolution guided NAS would be more simple and efficient for MSVS.

\vspace{-4mm}

\section{The MRP-CNN} \label{sec:method}

 \subsection{Proposed framework}\label{proposedframe}
The proposed MRP-CNN in Fig.~\ref{Block}(a)  is composed of five stacked Blocks\footnote{The number of blocks was choosen empirically here, and can be chosen flexibly in an application. Using more (than 5) stacked blocks, a higher separation performance may be obtained, but with a computationally higher complexity.}. Each Block  (indexed by $i$, $1$$\leq$$i$$\leq 5$) works as a basic unit to extract multi-resolution features and five Blocks form  a stacked structure.  Skip connections (dotted lines in Fig.~\ref{Block}(a)) can be optionally used between different Blocks to  improve the separation performance. 

As  illustrated in Fig.~\ref{Block}(b), each Block   consists of a convolution-group (CG), multiple pooling layers (PLs,   indexed by $j$, $1$$\leq$$j$$\leq$$J$), concatenation, and post-convolution-group (PCG) layer. Skip connection can be  used optionally (dotted lines in Fig.~\ref{Block}(b)). The $j$-th PL  in the $i$-th Block is composed with three components: an average  pooling operator   of size $T_{i,j}$$\times$$F_{i,j}$, a PCG layer,  and an upsampling operation. Each pooling layer (PL, $1$$\leq$$j$$\leq$$J$) is responsible for extracting  one specific resolution feature and the  Block which has multiple PLs can extract multi-resolution features. The CG  and PCG in each Block have the same structure. As shown in Fig.~\ref{Block}(c), both CG  and PCG are made of two consecutive convolution layers with the same size of $3\times$$3$$\times$$C$ and a possible skip connection, where $3$$\times$$3$ represents the kernel size of 2D convolutional operator and $C$ is the channel number.

Using the hyper-parameters (e.g., $T_{i,j}$, $F_{i,j}$, $C$, etc.) and flexible components (e.g., skip connection) of the  basic MRP-CNN framework,  many different  MRP-CNN architectures can be induced. For example, in each Block, the exact  PL number, i.e., $J$, can be adjusted by the data-driven evolution process of E-MRP-CNN. In particular, when the size of the average pooling operator of one PL is changed to $T_{i,j}$$=$$F_{i,j}$$=$$1$ during  the evolution process, this PL  will not be used in the current Block. In addition, the CG/PCG can have different channel numbers (different $C$) and when $C$$=$$0$, CG/PCG is turned into direct connection;  skip connections can be used optionally between different Blocks;  nonlinear activation functions can be different (e.g., ReLU or sigmoid).  Hence, the proposed MRP-CNN provides a flexible framework for MSVS.

 \vspace{-2mm} 
 \subsection{Encoding method}\label{encoding method}
The encoding process is to assign each specific  MRP-CNN architecture a unique code, i.e., the  gene. With the gene-encoded MRP-CNN architectures, a search space is constructed, thus enabling our NAS to find the appropriate architectures for MSVS  (see Section \ref{sec:method2}) under the defined objective. For the convenience of presentation, we divide the proposed MRP-CNN framework in Fig.~\ref{Block} into the following four levels from low to high 
$$ \text{\emph{Convolution-level}} \subset \text{\emph{Pooling-level}} \subset \text{\emph{Block-level}} \subset \text{\emph{Full-level}},$$
where \emph{Convolution-level} represents convolutional layers and CG and PCG belong  to this level, the \emph{Pooling-level}, \emph{Block-level}, and \emph{Full-level} correspond to PL, Block, and the whole MRP-CNN structure, respectively. The whole MRP-CNN structure can be encoded as in Table~\ref{encodingscheme}, where all the four levels are included.

\begin{table}[!t]
\vspace{-4mm}
\caption{Encoding method of the proposed MRP-CNN.} \label{tab:geneencoding} 
\vspace{-2mm}
\center
  \setlength
  {\tabcolsep}{0.6em}
\renewcommand\arraystretch{0.95}
\begin{tabular}{|l| l| l| l| l| }
\hline
\multicolumn{1}{|c|}{$\mathrm{FC}$} & \multicolumn{4}{c|}{2bit: 00(32), 01(64), 11(128), 10(256)} \\

\hline
\multicolumn{1}{|c|}{$\mathrm{FS}$} &\multicolumn{4}{c|}{10bit: b-bb-bbb-bbbb (b $\in \{0, 1\}$)}\\ 
\hline
\multirow{17}{*}{$\mathrm{B}$} & \multirow{5}{*}{$\mathrm{CG}$} & \multicolumn{2}{c|}{$\mathrm{C}$} & 2bit: 00(None), 01(32), 11(64), 10(128) \\
&& \multicolumn{2}{c|}{$\mathrm{S}$} & 1bit: 0(No), 1(Yes)\\
&& \multicolumn{2}{c|}{$\mathrm{A}$} & 1bit: 0(ReLU), 1(Sigmoid)\\
&& \multicolumn{2}{c|}{$\mathrm{A}$} & 1bit: 0(ReLU), 1(Sigmoid)\\
\cline{2-5}
& \multirow{5}{*}{$\mathrm{PL}$} & \multicolumn{2}{l|}{$\mathrm{PS}$} & (2bit)x(2bit): 00(1), 01(4), 11(16), 10(64) \\
\cline{3-5}
&& \multicolumn{2}{l|}{$\mathrm{PC}$} & 2bit: 00(16), 01(32), 11(64), 10(128) \\
\cline{3-5}
&& \multirow{3}{*}{$\mathrm{PCG}$} & $\mathrm{S}$ & 1bit: 0(No), 1(Yes) \\
&&& $\mathrm{A}$ &1bit: 0(ReLU), 1(Sigmoid)\\
&&& $\mathrm{A}$ & 1bit: 0(ReLU), 1(Sigmoid)\\
\cline{2-5}
&$\mathrm{PL}$ &\multicolumn{3}{c|}{....}\\
\cline{2-5}
&$\mathrm{....}$ &\multicolumn{3}{c|}{....}\\
\cline{2-5}
& \multirow{3}{*}{$\mathrm{PCG}$} & \multicolumn{2}{c|}{$\mathrm{S}$} & 1bit: 0(No), 1(Yes) \\
&& \multicolumn{2}{c|}{$\mathrm{A}$} & 1bit: 0(ReLU), 1(Sigmoid) \\
&& \multicolumn{2}{c|}{$\mathrm{A}$} & 1bit: 0(ReLU), 1(Sigmoid) \\

\hline
\multicolumn{1}{|c|}{....} & \multicolumn{4}{c|}{....} \\ 
\hline

\multicolumn{1}{|c|}{$\mathrm{B}$} & \multicolumn{4}{c|}{....} \\ 
\hline
\end{tabular}
\label{encodingscheme} 
\vspace{-3mm}
\end{table} 

\subsubsection{Full-level}\label{fullencode}
The \emph{Full-level}, i.e., the whole MRP-CNN structure, is encoded by $\mathrm{FC-FS-B-B-B-B-B},$ where $\mathrm{{FC}}$ encodes the  number of channels of the last PCG layer in all  Blocks, i.e., $C_{i,J+1}$ (see Fig.~\ref{Block}(b)), $\mathrm{{FS}}$  encodes  possible skip connections between different Blocks, $\mathrm{{B}}$ stands for Block, and ``$-$'' represents concatenation of codes.

The value of $\mathrm{{FC}}$ can be $32/64/128/256$, as  shown in Table~\ref{encodingscheme}, where we use  2 bits to represent four options: 00(32), 01(64), 11(128), 10(256), respectively.  
Here, the same $\mathrm{{FC}}$ (one of the four options) is used for all Blocks in one MRP-CNN structure,  since  the output channels of different Blocks should  be the same to enable skip connections.
 
The  $\mathrm{{FS}}$ is encoded in form of  ``b-bb-bbb-bbbb" using ten bits (see the second row in Table~\ref{encodingscheme}). The first bit `b' stands for   the skip connection from the first Block to the second Block, the second `bb'  stands for    skip connections from the first and  second Block to the third Block, and so on.  The value of  b decides if  skip connection exists (b=$1$) or not (b=$0$). 

\subsubsection{Block-level} 
This level is important to extract multi-resolution features. Each Block  is encoded as 
$$ \mathrm{B}=\mathrm{CG-\underbrace{\rm{PL}-\cdots-PL}_{\textit{J}} -PCG},$$where $\mathrm{CG}$,  $\mathrm{PL}$, and $\mathrm{PCG}$ have been defined earlier. Both CG and  PCG belong to \emph{Convolution-level} and PLs working in parallel belong to \emph{Pooling-level}.

\begin{table}[t]
\vspace{-4mm}
\captionof{table}{The code (gene) of an example MRP-CNN.}
\vspace{-2mm}
\center
  \setlength
  {\tabcolsep}{0.1em}
\renewcommand\arraystretch{0.95}
\begin{tabular}{|l| l| l| l|  l |l | l | l | }
\hline
\multicolumn{1}{|c|}{$\mathrm{FC}$} & \multicolumn{7}{c|}{11(128)}\\ \hline 
\multicolumn{1}{|c|}{$\mathrm{FS}$} & \multicolumn{7}{c|}{0-00-000-0000}\\ \hline 
\multicolumn{3}{|c|}{Blocks $\to$} & Block  1  & Block 2   & Block 3   & Block 4   & Block 5 \\ \hline 
  \multirow{4}{*}{$\mathrm{CG}$} & \multicolumn{2}{c|}{$\mathrm{C}$} & 11(64) & 11(64) & 11(64) & 11(64) & 11(64) \\
 & \multicolumn{2}{c|}{$\mathrm{S}$} & 1(Yes) & 1(Yes) & 1(Yes) & 1(Yes) & 1(Yes) \\
 & \multicolumn{2}{c|}{$\mathrm{A}$} & 0(ReLU) & 0(ReLU) & 0(ReLU) & 0(ReLU) & 0(ReLU) \\
 & \multicolumn{2}{c|}{$\mathrm{A}$} & 0(ReLU) & 0(ReLU) & 0(ReLU) & 0(ReLU) & 0(ReLU) \\
\hline
 \multirow{5}{*}{$\mathrm{PL}$} & \multicolumn{2}{l|}{$\mathrm{PS}$} & 0011(1x16) & 0011(1x16) & 0011(1x16) & 0011(1x16) & 0011(1x16) \\
\cline{2-8}
 & \multicolumn{2}{l|}{$\mathrm{PC}$} & 11(64) & 11(64) & 11(64) & 11(64) & 11(64) \\
\cline{2-8}
 & \multirow{3}{*}{$\mathrm{PCG}$} &  $\mathrm{S}$ & 1(Yes) & 1(Yes) & 1(Yes) & 1(Yes) & 1(Yes) \\
 && $\mathrm{A}$ & 0(ReLU) & 0(ReLU) & 0(ReLU) & 0(ReLU) & 0(ReLU) \\
 && $\mathrm{A}$ & 0(ReLU) & 0(ReLU) & 0(ReLU) & 0(ReLU) & 0(ReLU) \\
\hline
  \multirow{5}{*}{$\mathrm{PL}$} & \multicolumn{2}{l|}{$\mathrm{PS}$} & 0000(1x1) & 0000(1x1) & 0000(1x1) & 0000(1x1) & 0000(1x1) \\
\cline{2-8}
 & \multicolumn{2}{l|}{$\mathrm{PC}$} & \textcolor{lightgray}{11(64)} & \textcolor{lightgray}{11(64)} & \textcolor{lightgray}{11(64)} & \textcolor{lightgray}{11(64)} & \textcolor{lightgray}{11(64)} \\
\cline{2-8}
& \multirow{3}{*}{$\mathrm{PCG}$} & $\mathrm{S}$ & \textcolor{lightgray}{1(Yes)} & \textcolor{lightgray}{1(Yes)} & \textcolor{lightgray}{1(Yes)} & \textcolor{lightgray}{1(Yes)} & \textcolor{lightgray}{1(Yes)} \\
 && $\mathrm{A}$ & \textcolor{lightgray}{0(ReLU)} & \textcolor{lightgray}{0(ReLU)} & \textcolor{lightgray}{0(ReLU)} & \textcolor{lightgray}{0(ReLU)} & \textcolor{lightgray}{0(ReLU)} \\
 && $\mathrm{A}$ & \textcolor{lightgray}{0(ReLU)} & \textcolor{lightgray}{0(ReLU)} & \textcolor{lightgray}{0(ReLU)} & \textcolor{lightgray}{0(ReLU)} & \textcolor{lightgray}{0(ReLU)} \\
\hline
  \multirow{4}{*}{$\mathrm{PCG}$} & \multicolumn{2}{c|}{$\mathrm{S}$} & 1(Yes) & 1(Yes) & 1(Yes) & 1(Yes) & 1(Yes) \\
 & \multicolumn{2}{c|}{$\mathrm{A}$} & 0(ReLU) & 0(ReLU) & 0(ReLU) & 0(ReLU) & 0(ReLU) \\
 & \multicolumn{2}{c|}{$\mathrm{A}$} & 0(ReLU) & 0(ReLU) & 0(ReLU) & 0(ReLU) & 0(ReLU) \\
\hline
 \end{tabular}
\label{tab:gene_Hand} 
\vspace{-4mm}
\end{table}

\subsubsection{Convolution-level}\label{convencoding}
The CG and PCG  which  have the same architecture (see Fig.~\ref{Block}(c)) are encoded differently. The $\mathrm{CG}$ is encoded as $$\mathrm{CG=C-S-A-A,}$$ where $\mathrm{C}$ encodes the number of channels of  convolutional layers in CG, i.e., $C_{i,0}$ in Fig.~\ref{Block}(b), $\mathrm{S}$ stands  for the skip connection ($\mathrm{S}$ $\in \{0, 1\}$), and two consecutive bits $\mathrm{A-A}$ imply the activation functions for the two-layer convolution operators,  where  $\mathrm{A}$$=$$0$ represents  ReLU and $\mathrm{A}$$=$$1$ represents  Sigmoid. The values of $\mathrm{C}$ can be $0/32/64/128$. When  $\mathrm{C}$$=$$0$, the CG turns into a direct connection, i.e.,  there is no convolution, activation, or skip connection. In this case, the $\mathrm{S-A-A }$ is ignored.

The code of PCG is  similar to CG but without  the channel number information, i.e.,
$$\mathrm{PCG=S-A-A}.$$
According to  Fig.~\ref{Block}(b), the PCG is employed in both Block  and  PL. Thus  the   channel number of  PCG in Block and in PL is decided by $\mathrm{FC}$ in  \emph{Full-level} and $\mathrm{PC}$  in  \emph{Pooling-level} (see the following), respectively.

\subsubsection{Pooling-level} Each PL is encoded using
$$\mathrm{PL=PS-PC-PCG},$$
where $\mathrm{PS}$ is the size of  pooling operator in PL, $\mathrm{PC}$ is the channel number of  PCG (i.e., $C_{i,j}$ of the  $j$-th PL  in the $i$-th Block  in Fig.~\ref{Block}(b)), and $\mathrm{PCG}$ represents the post convolution group. For the $j$-th PL in the $i$-th Block,  $\mathrm{PS}$ is defined as [$T_{i,j}$,$F_{i,j}$], where $T_{i,j}$ is the downsampling size in time axis and $F_{i,j}$ in frequency axis. 
When $T_{i,j}=F_{i,j}=1$, the $j$-th PL will not appear in the $i$-th  Block and  the  code $\mathrm{PC-PCG}$  will be ignored. We use 2 bits to encode $T_{i,j}$ and $F_{i,j}$ of $\mathrm{PS}$, respectively. As shown in Table~\ref{encodingscheme},  four possible values  are  represented by 00(1), 01(4), 11(16), and 10(64). The $\mathrm{PC}$ is also encoded  by 2 bits: 00(16), 01(32), 11(64), and 10(128). The upsampling operator in PL is not encoded, since it has no freedom but to upsample  the extracted features back to the same size as the input of the current PL.

A simple  example of MRP-CNN is shown in Table \ref{tab:gene_Hand}, where  all five Blocks have two PLs. The $\mathrm{PS}$ of the second PL is 0000 ($T_{i,2}$$=$$F_{i,2}$$=$$1$), i.e., the  $\mathrm{PC}$ and $\mathrm{PCG}$  are ignored (shown  in gray).  This  MRP-CNN (or other MRP-CNN architectures)  can be used as a seed in E-MRP-CNN.

\vspace{-2mm}
\section{The E-MRP-CNN}\label{sec:method2}
Using  the above  encoding method, each possible MRP-CNN structure can be represented by a unique code (i.e., gene). All these genes form a big searching space. The proposed E-MRP-CNN utilizes genetic algorithm to  automatically search  effective   genes, i.e.,  effective MRP-CNN structures, from this searching space. Here, we propose two types of evolution schemes: the single-objective and the multi-objective  E-MRP-CNN scheme.

Both single/multi-objective schemes  start with an initial population, which is  made of  a seed gene (a specific MRP-CNN  structure) and other genes (structures) randomly mutated from this seed gene. After initialization, the single/multi-objective schemes iteratively generate  new offspring genes  by applying genetic operations (i.e., crossover and mutation) to randomly selected gene(s) from the  current population. The new offspring genes are decoded to corresponding  MRP-CNN structures which are then trained/tested and assigned with  fitness values. The fitness values for single-objective  and  multi-objective  schemes are  computed  in different ways: the single-objective scheme considers  only the separation performance  while  the multi-objective scheme  considers both separation performance and model complexity. When the fitness values for all genes are computed, the genes with low fitness in current generation will be  removed. This evolution iteration is   repeated and finally ended in a generation made of well-performing genes (structures). 

\vspace{-2mm}
\subsection{Single-objective E-MRP-CNN }\label{singleobj}

According to BSS-EVAL toolkit \cite{vincent2006performance}, there are usually three metrics to measure the separation performance of MSVS: source-to-distortion ratio (SDR), source-to-interferences ratio (SIR), and sources-to-artifacts ratio (SAR). As a proof of concept, we choose SDR as the  fitness function to guide the evolution process of the single-objective scheme, because  it is a global performance measure considering three goals\footnote{According to~\cite{vincent2006performance}, three goals are (i) rejection of the interferences, (ii) absence of forbidden distortions and ``burbling'' artifacts, and (iii) rejection of the sensor noise.} as equally  important~\cite{vincent2006performance}. In particular,  since   each gene is only partially trained in the evolution process (to accelerate the computation), the global  measure   SDR would be more suitable than the SIR and SAR.

The single-objective scheme  is presented in Algorithm~\ref{SingleObj-E-MR-CNN}, where Rows 1-4 show the  initialization process and Rows 5-12 show the  evolution process. 

\subsubsection{Initialization process}
\begin{itemize}
\item In the  first step (Row 1), we generate the initial population of size $n$,  including one seed gene and  the other $n-1$ genes  randomly mutated from this seed. To do this, the  $n_b$ bits of the seed gene are flipped to generate a new gene, where $n_b$  is a random number and  $1\leq n_b\leq u$ ($u$ is the maximum flipping number). We repeat this process until $n-1$ different   genes are obtained.
\item In the second step (Row 2), we divide the training dataset denoted by $\mathscr{D}$ into three subsets  $\mathscr{D}\to\{\mathscr{D}_{tr}, \mathscr{D}_{te}, \mathscr{D}_{v}\}$, where the  training subset $\mathscr{D}_{tr}$ is used for training, the testing subset $\mathscr{D}_{te}$ is used for computing  the fitness, and the validation subset $\mathscr{D}_{v}$  is used to decide when to stop the evolution process of the single-objective scheme. 
\item In the third step (Row 3-4), we compute the fitness  of each gene in the  initial population. Specifically, the MRP-CNN structure  decoded from each gene is trained with $\mathscr{D}_{tr}$ for only a few iterations (i.e., partial training). These partially trained structures are  tested on  $\mathscr{D}_{te}$ and we compute the averaged SDR performance\footnote{This averaged SDR score is computed on the subset $\mathscr{D}_{te}$, which can be considered as an approximation of the separation performance on the full testing dataset in the final evaluation.}   over all clips of $\mathscr{D}_{te}$ as the  fitness of each gene. The genes with  low-fitness are removed according to the population limit $Z$. 
\end{itemize}

 \subsubsection{Evolution  process}
\begin{itemize}
\item In each iteration of evolution, we use crossover (Row 6) and mutation (Row 7) operators   to generate new offspring genes.  The crossover operator recombines the information of the two randomly selected genes, where  one gene is used as the baseline and  each bit within it has a probability (prob.) $p_1$ to be exchanged with the corresponding bit of the other gene. We apply crossover to create  $o_c$ new offsprings. The mutation operator produces  a  new offspring by randomly flipping each bit of one gene with a  prob. $p_2$. We apply the mutation operator to each gene of the current generation and the newly obtained $o_c$  offsprings (generated by the crossover) to create total $o_m=o_c+Z$ new offsprings.

\item The SDR fitness values  of all new offsprings ($o_c+o_m$)  are  computed  (Row 8). All   populations including the  new offsprings ($o_c+o_m$)  and the current populations ($Z$) are sorted by their fitnesses (Row 9) and the  low-fitness genes are removed according to the population limit  $Z$ (Row 10). 

\item We check if the stopping criterion is satisfied with the validation subset $\mathscr{D}_{v}$ (Row 11). Specifically, we test the best-fitness gene of the current generation on $\mathscr{D}_{v}$ to compute its SDR, which can be considered as the best SDR performance of the current generation.  This SDR is then compared with the SDRs of  several  recent generations and if  there is no improvement on  this value for a few generations ($S$ generation),  the evolution iteration will be stopped and 
the earliest generation with no SDR improvement will be given as the output. 
\end{itemize}

\begin{algorithm}[t]
\small
\caption {Single-objective E-MRP-CNN}
\begin{algorithmic}[1]

 \State Generate the initial population of size $n$
    \State Data preparation: training set $\mathscr{D}\to\{\mathscr{D}_{tr}, \mathscr{D}_{te}, \mathscr{D}_{v}\}$
  \State Compute SDR  fitness   of each gene in the  initial population
  \State Remove low-fitness genes according to population limit $Z$

  \For{$i =1$ to $N$ (maximum generation)}
    \State Generate $o_c$ new genes by crossover with prob. $p_1$
    \State Generate $o_m$ new genes by mutation with prob. $p_2$
    \State Compute SDR  fitness  for  all new offsprings

    \State Sort all genes (current+new) by SDR fitness
    \State Remove low-fitness genes by population limit $Z$
    \State break, if stopping criterion is satisfied 
  \EndFor

\end{algorithmic}
\label{SingleObj-E-MR-CNN}

\end{algorithm}

\vspace{-2mm}

\subsection{Multi-objective E-MRP-CNN}\label{multiobjective}
The single-objective scheme  evolves only  to improve  the  separation performance. Thus  it may pick up the more complex neural structures that provide better separation performance. Since the model complexity is an important factor for  limited  memory applications~\cite{GuoZYWY18}, the  multi-objective scheme is designed  to balance two objectives, i.e., model complexity and  separation performance. In fact, these two objectives are conflicting: a complicated structure is more likely to provide a higher performance than a simple one. Thus  the multi-objective scheme tries to  approximate  the Pareto front set, where many solutions are included and each solution provides a good separation performance under a fixed model complexity.

There are generally two  properties to design evolutionary multi-objective optimization algorithms:  convergence and diversity \cite{LiKZD15}. The convergence measures the  distances of solutions toward the Pareto front (i.e., Pareto-optimal front) which  should be as small as possible \cite{LiKZD15}. The diversity is the spread of solutions along the Pareto front and should be as uniform as possible \cite{LiKZD15}. For MSVS, the convergence encourages each  evolved structure to offer a  separation performance as good as possible under a certain complexity; the diversity encourages  the evolved  structures to be  various enough to handle different  complexity levels. To achieve these, the proposed multi-objective scheme is implemented based on NSGA-II~\cite{DebAPM02}, where the fast non-dominated sorting is used to promote convergence and the crowded-comparison operator is employed to address diversity~\cite{DebAPM02}.

The multi-objective scheme   is presented in Algorithm~\ref{algoNSGA-II}, where Rows 1-4 show the  initialization process and Rows 5-11 show the  evolution iteration. 

The first two steps  in the initialization process (Rows 1-2) are the same as  those in the single-objective scheme (note that the subset $\mathscr{D}_{v}$ is not used here).  In the third step, we compute the fitness of each gene  in the initial population, but instead of considering the SDR as the only fitness, we calculate both the SDR score  and the  model complexity (measured by the amount of parameters (Params)) of each gene. Then we use the  fast non-dominated sorting of NSGA-II~\cite{DebAPM02} to calculate the non-dominated levels of all  genes. 
By sorting all these levels with   crowded-comparison operator,  low-fitness genes are  removed according to the population limit $Z$ (Row 4).  

In  each iteration of the evolution, we use crossover (Row 6) and mutation (Row 7)   to generate $o_c$  and $o_m$ ($o_m=o_c+Z$) new offsprings, respectively. The SDR and model complexity of  all $o_c+o_m$ new offsprings are computed. Both the current populations ($Z$) and the new offsprings ($o_c+o_m$) are sorted by fast non-dominated sorting and crowded-comparison operator of NSGA-II. We  remove   low-fitness genes according to the population limit $Z$. The multi-objective scheme  will stop until the  maximum iteration number  is reached.

\begin{algorithm}[t]
\small
 \caption{Multi-objective E-MRP-CNN}
 \begin{algorithmic}[1] 
  \renewcommand{\algorithmicrequire}{\textbf{Input:}}
 \renewcommand{\algorithmicensure}{\textbf{Output:}}
  \State Generate the initial population of size $n$
   \State Data preparation: training set $\mathscr{D}\to\{\mathscr{D}_{tr}, \mathscr{D}_{te}, \mathscr{D}_{v}\}$
  \State  Compute SDR  and model complexity   and then perform fast non-dominated sorting and  crowded-comparison
  \State  Remove low-fitness genes according to the population limit $Z$
  \For {$i =1$ to $N$ (maximum generation)}
    \State Generate $o_c$ new genes by crossover with prob. $p_1$
    \State Generate $o_m$ new genes by mutation with prob. $p_2$
 \State Compute SDR  and Params for all new offsprings
 \State Sort all genes (current+new) using fast non-dominated sorting 

and crowded-comparison
 \State Remove low-fitness genes  by the population limit $Z$
\EndFor  
 \end{algorithmic} 
 \label{algoNSGA-II}
 \end{algorithm}

\section{Experiment setting}\label{sec:exp1}
\subsection{Datasets and evaluation metrics}\label{datasets}

The proposed method was evaluated on two popular datasets: MIR-1K~\cite{MIR} and DSD100~\cite{DSD100}. The MIR-1K dataset contains a thousand song clips extracted from
110 karaoke songs. For a fair comparison, we followed the evaluation conditions in~\cite{huang2015joint,yang2013low,SHN, weitao1}: $175$ clips performed by one male singer `abjones' and one female singer `amy' were used for  training, the other $825$ clips performed by $17$ singers were used for testing. On DSD100, songs of the "Dev" subset  were used  for training and  we followed \cite{SHN,  weitao1} to convert all sources to monophonic and then added  three sources except for the vocals  together to form the musical component  (i.e., the Acc) source.

The separation performance was  quantitatively measured by the BSS-EVAL toolkit~\cite{vincent2006performance} with respect to three criteria:  SDR, SIR, and SAR. Normalized SDR (NSDR)~\cite{ozerov2007adaptation} was calculated to show the improvement of SDR compared to the original mixture. 
Global NSDR (GNSDR), Global SIR (GSIR), and Global SAR (GSAR) were computed by taking the weighted means of the NSDRs, SIRs, SARs, respectively, over all the test clips weighted by their length \cite{huang2015joint, SHN}. Some qualitative results were also presented to verify the separation performance of the proposed method.

\vspace{-2mm}
\subsection{T-F masking framework }\label{tfmasking}

\begin{figure}[t]
\vspace{-6mm}
\center
\includegraphics[width=0.4\textwidth]{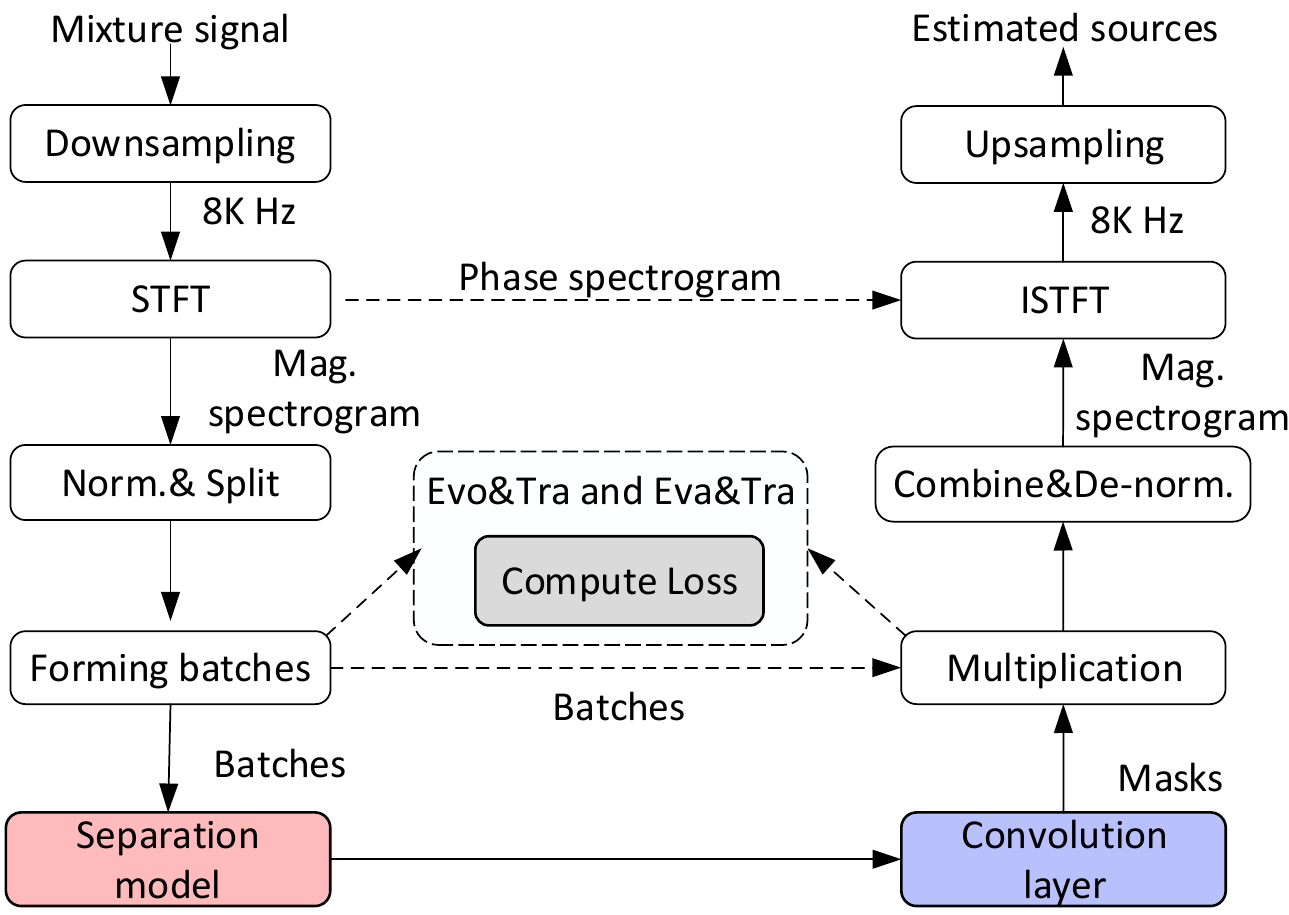} 
\caption{The T-F masking framework.}
\label{TFframework}
\vspace{-4mm}
\end{figure}

The proposed MRP-CNN and E-MRP-CNN were evaluated based on the  T-F masking framework  in Fig. \ref{TFframework}, where the red rectangular is the key separation module\footnote{Although it is advantageous to  use   independent separation  module for each source, i.e.,  two separation modules for two sources, it is computationally expensive according to \cite{SHN}. Hence,   following \cite{SHN}, we use only one separation module.} (can be the proposed structure or other compared structures). The output of the separation module is fed to the convolution layer (blue rectangular), which has two outputs for estimating  the T-F masks  for  Vocal and Acc sources  in MSVS. This  framework (or similar frameworks) is widely employed in many MSVS/MSS methods (see~\cite{SHN}, \cite{MimilakisDVS17,MimilakisDSSVB18,LiutkusB15}).

The above  framework was used  in both evolution  process (denoted by Evo)  and the final evaluation (denoted by Eva). 
For each situation, we have two  scenarios: training (Tra) and testing (Tes). For Evo, we trained the evolved structures in the T-F masking framework  using $\mathscr{D}_{tr}$ (Evo\&Tra) and then tested the trained structures on $\mathscr{D}_{te}$ (Evo\&Tes) to obtain the SDR fitness. For Eva, the final evolved structures were  trained in the T-F masking framework using the full training set (Evo\&Tra) and then tested on the full testing set (Eva\&Tes). 

%

When using the T-F masking framework, the input mixture signal (time-domain)  was first  downsampled to $8$ kHz  to speed up computation~\cite{SHN}. The $8$ kHz mixture signal was transformed to its spectrogram (a complex matrix) via  short-time Fourier transform (STFT) using a window size of $1024$ and a hop size of $256$.  The  magnitude spectrogram of the mixture was  normalized by dividing  its maximum value and  then split into blocks of size $512\times64$ (frequency$\times$frames) to form batches. The  batches of the mixture were fed to the separation  module and its output was fed to the convolution layer to predict the  masks (in batches) for Vocal and Acc sources. The predicted masks were used in (\textit{i}) the  training process (Evo\&Tra and Eva\&Tra) to  compute  the loss function and (\textit{ii}) the testing process (Evo\&Tes and Eva\&Tes) to output  the  time-domain estimated sources. 

 In the training process (Evo\&Tra  and Eva\&Tra), the loss function $L_{1,1}$ norm in~\cite{ronneberger2015u, SHN} was adopted for a fair comparison. 	 Formally, given the  mixture  $\mathbf{X}$, the $i$-th ground truth  source $\mathbf{Y}_i$, and the predicted mask $\mathbf{M}_i$ for the $i$-th source ($i=1...s$, $s=2$ in MSVS), the loss function is defined as 
$\mathcal{J} = \sum_{i=1}^{s}   \| \mathbf{Y}_i - \mathbf{X} \odot \mathbf{{M}}_{i} \|_{1,1},$ where $\odot$ denotes the element-wise multiplication of  matrices. Note that when computing the loss funciton, the magnitude spectrograms of  the ground-truth Vocal and Acc  sources  were also normalized by dividing the maximum value  of  their  mixture's magnitude spectrogram.

 In testing process (Evo\&Tes  and Eva\&Tes), the predicted masks for Vocal and Acc were truncated to the range of $[0.0, 1.0]$ and  multiplied   with the normalized spectrogram of the mixture~\cite{SHN}.  After de-normalization and batch combination, the  time-domain sources were obtained via inverse STFT (ISTFT) followed by upsampling.

In particular, for Eva\&Tra, two  data augmentation operations, gain and sliding, were applied to original time-domain  ground-truth sources, to creat new mixtures. The gain operation multiplied the original source by a random factor $a$ ($0.5$$\leq$$a$$\leq$$1.5$) and the sliding operation added a random short delay $d$ ($0$s$\leq$$d$$\leq$$0.5$s)  to the beginning of  the original source. The newly obtained  ground-truth sources were mixed to form new mixtures. The ratio of the augmented data   to the original data   is $1$$:$$4$.  All the differences of using the T-F masking framework for four scenarios are summarized in Table~\ref{differescenarios}.

\begin{table} [t]
\vspace{-4mm}
\captionof{table}{Differences scenarios of using the T-F framework.} \label{differescenarios} 
\begin{center}
\center
\setlength
{\tabcolsep}{0.5em}
\begin{tabular}{|c|c|c|c|c|}
\hline
Scenarios & Evo\&Tra & Evo\&Tes & Eva\&Tra& Eva\&Tes  \\
\hline
 
Data augmentation & & & $\checkmark$ & \\    \hline       
Training dataset $\mathscr{D}$ &$\checkmark$  &$\checkmark$ &$\checkmark$  &\\    \hline
Testing dataset & & &  &$\checkmark$\\   \hline
Subset $\mathscr{D}_{tr}$ of $\mathscr{D}$ &$\checkmark$ & & & \\   \hline 
Subset $\mathscr{D}_{te}$ of $\mathscr{D}$  & & $\checkmark$ & & \\   \hline
Subset $\mathscr{D}_{v}$  of $\mathscr{D}$  & &$\checkmark$ (Single) &  & \\     \hline
Truncation &  & $\checkmark$ &   & $\checkmark$  \\      \hline  
\end{tabular}
\end{center}

\end{table}
 
\vspace{-2mm}
 \subsection{Hyperparameters of the E-MRP-CNN}\label{HperParEvolution}
Table~\ref{tab:evocfg} lists the hyperparameters of the E-MRP-CNN. 
Since the multi-objective scheme  requires more diversity,  its  population limit $Z$ and mutation number $o_m$ were higher than those of the single-objective scheme. 
 For MIR-1K, the $\mathscr{D}_{tr},\mathscr{D}_{te}$, and $\mathscr{D}_{v}$ were set as $100/55/20$ (clips). For DSD-100, the $\mathscr{D}_{tr},\mathscr{D}_{te}$, and $\mathscr{D}_{v}$ were set as   $30/15/5$ (songs). For the multi-objective scheme, $\mathscr{D}_{v}$ and $S$  were not used. 

\begin{table} [t]
\caption{Hyperparameters of the E-MRP-CNN.}
\center
\setlength
{\tabcolsep}{0.4em}
\begin{tabular}{ccccccccccccc}
\hline
Scheme &$n$ & $u$ & $N$ & $Z$ & $o_c$ & $o_m$ & $p_1$ & $p_2$& $\mathscr{D}_{tr}$ & $\mathscr{D}_{te}$ & $\mathscr{D}_{v}$ & $S$\\
\hline
Single &22&20&100& 15 & 10 & 25 & 0.5 & 0.02  &         100/30                  &     55/15		& 	20/5   & 8   \\
Multi.&37&20&100& 25 & 10 & 35 & 0.5 & 0.02    &   100/30           &       55/15		 &		--  & --  \\
 \hline
\end{tabular}
\label{tab:evocfg}
\vspace{-2mm}
\end{table}

\begin{figure*}[t]

\vspace{-6mm}
\center

\subfigure[Multi-objective scheme]{
\includegraphics[width=0.48\textwidth]{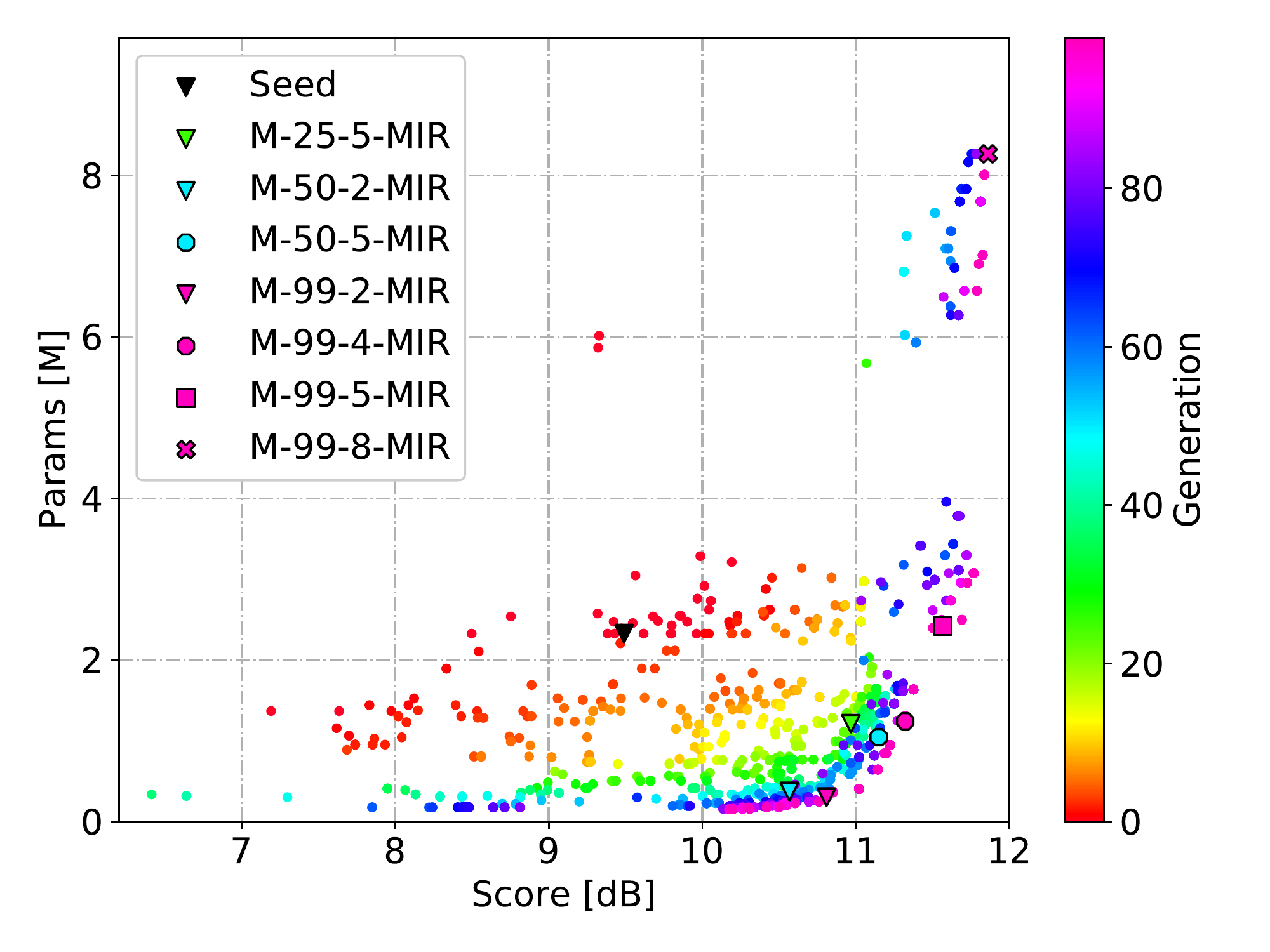} 
\label{fig_mir1k_nsga2}}
\subfigure[Single-objective scheme]{
\includegraphics[width=0.48\textwidth]{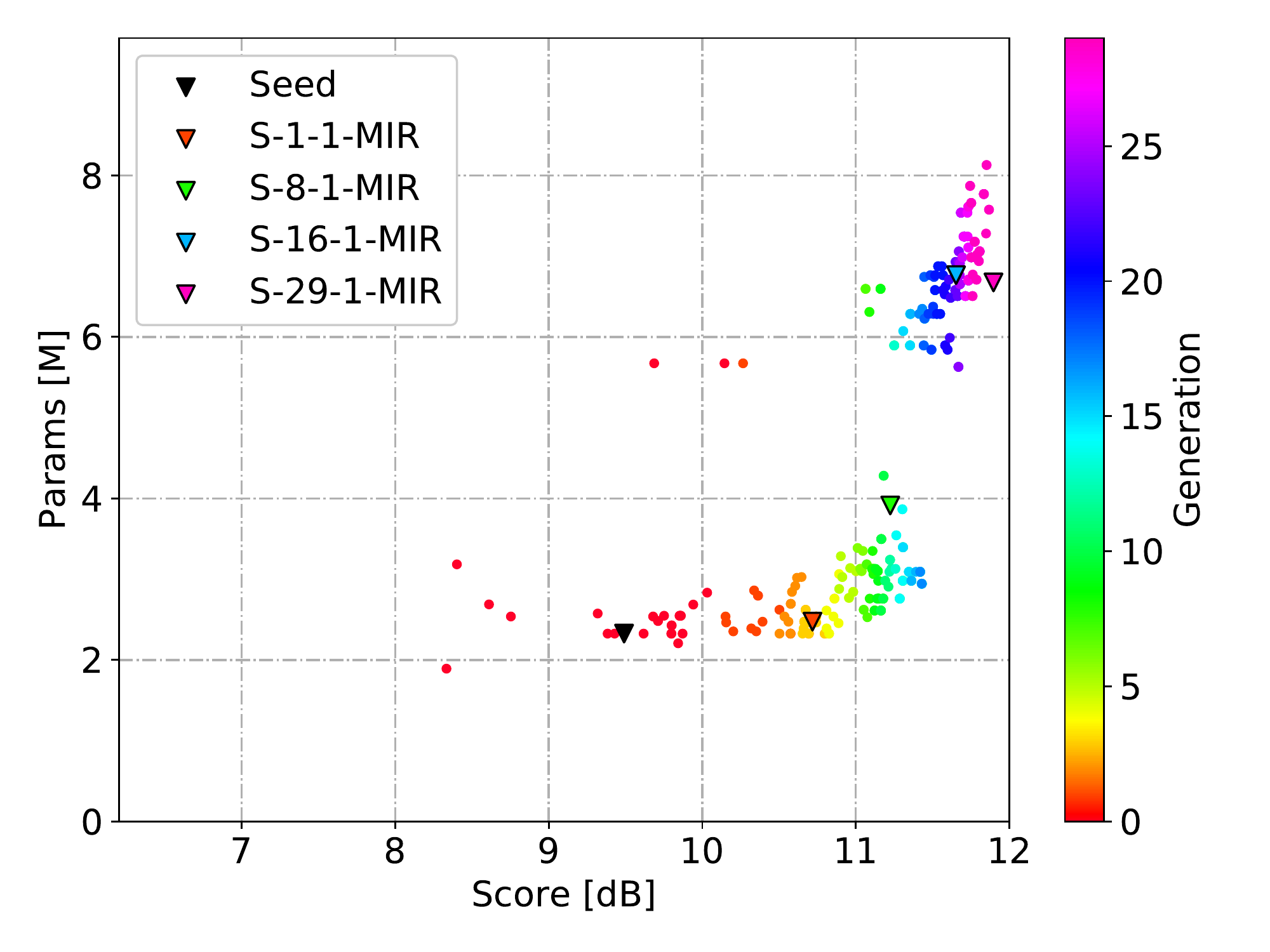} 
\label{fig_mir1k_single}}

\vspace{-2mm}

\caption{The evolution processes of  the single-objective  and multi-objective E-MPR-CNN on MIR-1K.}
\label{mir1kgen}
\vspace{-4mm}
\end{figure*}

\begin{figure*}[t]
\center

\subfigure[Multi-objective scheme]{
\includegraphics[width=0.48\textwidth]{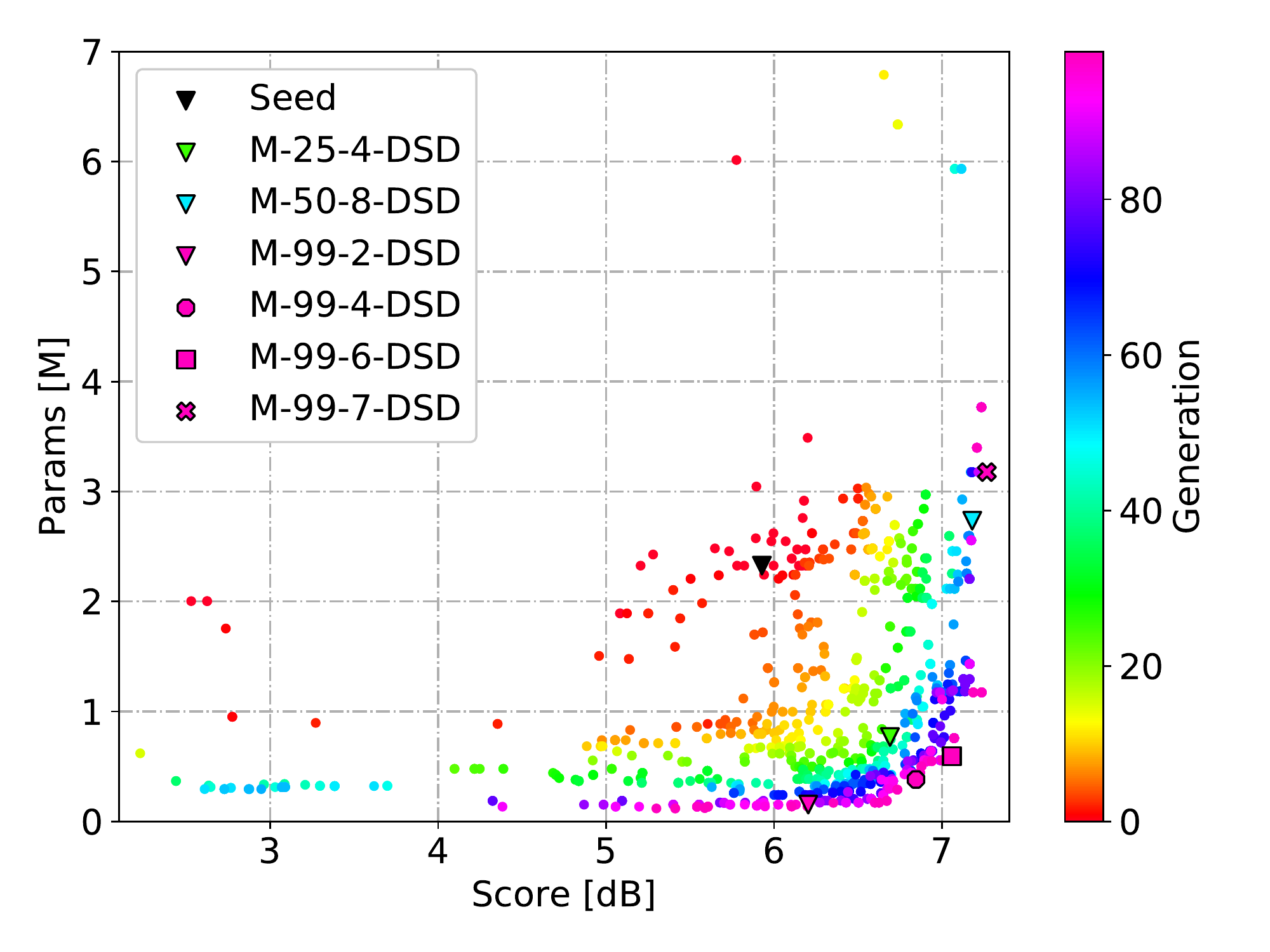} 
\label{fig_dsd2_nsga2}}
\subfigure[Single-objective scheme]{
\includegraphics[width=0.48\textwidth]{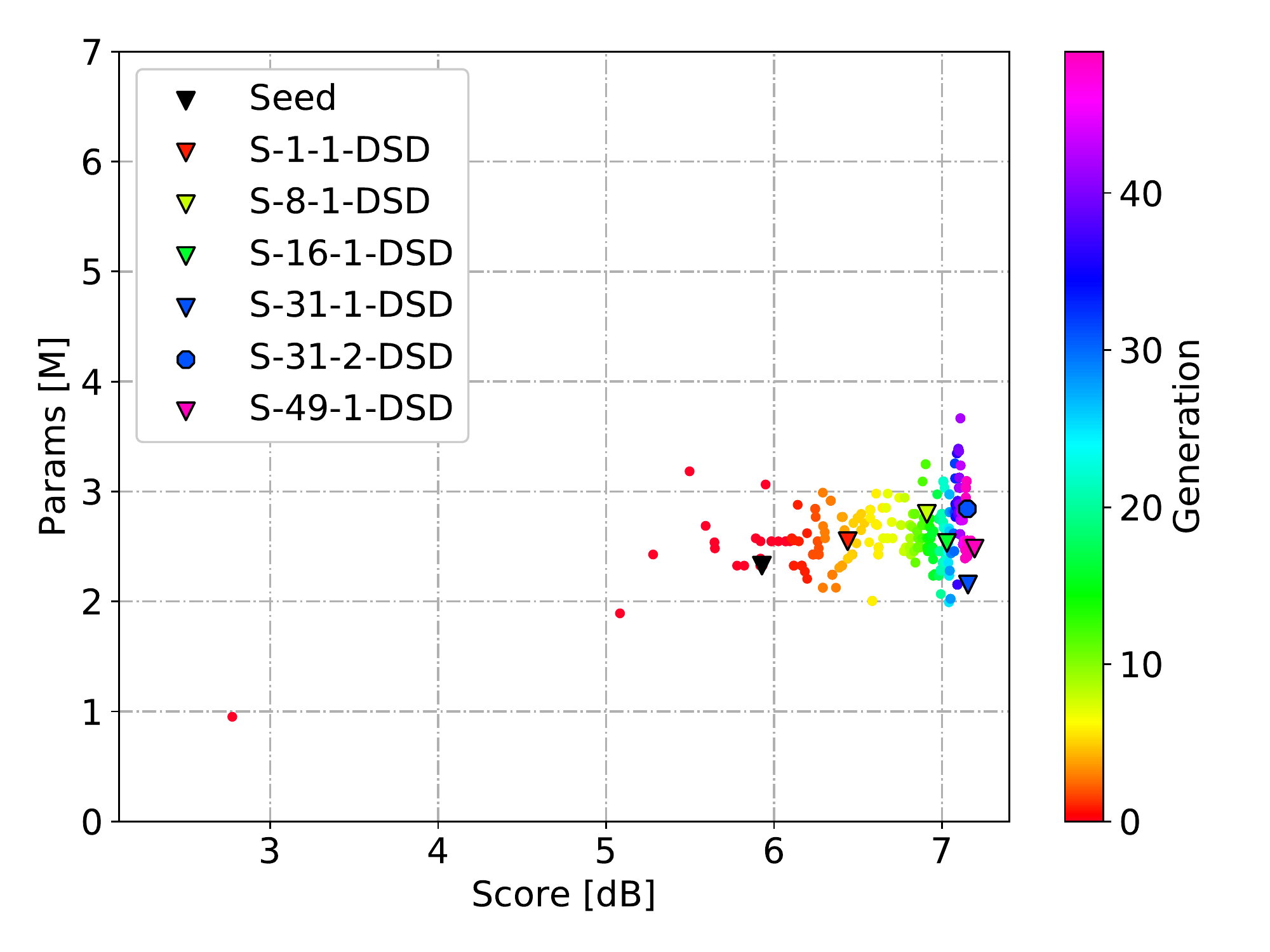} 
\label{fig_dsd2_single}}

\vspace{-2mm}

\caption{The evolution processes of  the single-objective and multi-objective   E-MPR-CNN on DSD100.}
\label{DSD100generations}
\vspace{-4mm}
\end{figure*}

\vspace{-2mm}
\subsection{Training parameters}\label{trainingparameters}
The Adam optimizer~\cite{kingma2014adam} was employed to train the T-F masking framework. In Evo\&Tra, we aim to compute the `fitness' of each gene, so  the T-F masking framework was only partially trained with $1,500$ iterations for the MIR-1K dataset and $3,100$ iterations for the DSD100 dataset using  batch size $2$. In  Eva\&Tra, the  framework was fully trained with $63,000$ iterations for the MIR-1K dataset and $630,000$ iterations for the DSD100 dataset using  batch size $3$.


In both Evo\&Tra  and Eva\&Tra,  two tricks were used: (i) cosine decay learning rate and warm restart \cite{LoshchilovH17} and (ii) learning rate warmup \cite{GoyalDGNWKTJH17}. 
For (i), we set $T_0$$=$$100$ and $T_{m}$$=$$2$ for  both datasets in Evo\&Tra,   and $T_0$$=$$1000$ ($10000$) for MIR-1K (DSD100) and  $T_{m}$$=$$2$ in Eva\&Tra, where $T_0$ is the length of first decay period \cite{LoshchilovH17} and $T_{m}$ is the multiplication factor for decay period length at every new warm restart  \cite{LoshchilovH17}. The maximum learning rate for Evo\&Tra and Eva\&Tra was $3\times10^{-4}$.  The minimum learning rates  for Evo\&Tra and Eva\&Tra were $1\times10^{-4}$ and   $1\times10^{-5}$,  respectively (more details can be found in \cite{LoshchilovH17}).
For (ii), we scaled the learning rate in the first $100$ ($1000$) iterations for Evo\&Tra (Eva\&Tra) with a factor $0.3$, to avoid the maximum learning rate being too large for some genes.

\vspace{-2mm}

\section{Experimental results}\label{sec:exp2}


\subsection{Evolution process of the E-MPR-CNN} \label{evoprocess}

For both single-objective and  multi-objective schemes, the  MRP-CNN structure in Table~\ref{tab:gene_Hand} was used as the seed of the initial population of E-MRP-CNN on two datasets.  The evolved genes (structures) of E-MRP-CNN are represented in form of ``S/M-G-Index-Dataset'', where S and M denote the single-objective scheme and multi-objective  scheme, respectively, G represents the generation (evolution) number, Index is the   gene index in the G-th generation, and Dataset can be MIR (MIR-1K) or DSD (DSD100). For single-objective scheme, the Index is the SDR ranking of a gene in the current generation; for multi-objective scheme, the Index is the  gene index in the current generation. For example, ``S-25-2-MIR'' represents the structure with the second highest SDR performance  in the $25$th generation of the single-objective scheme on the  MIR-1K dataset,  ``M-99-2-DSD'' represents the No. $2$ evolved structure  in the $99$th generation of the multi-objective scheme on  the  DSD100 dataset.

We recorded the dynamic evolution process  of  E-MRP-CNN in Fig.~\ref{mir1kgen} for the MIR-1K dataset and Fig. \ref{DSD100generations} for the DSD100 dataset. The vertical axis in each figure  represents the model complexity  measured by  Params  and the horizontal axis represents the fitness score measured by  SDR. Each colored data point  stands for a gene, i.e., a MRP-CNN  structure. The genes of  different generations are distinguished by colors changing  from red  (initial generation) to pink (highest generation). We set the highest evolution number as $99$. In our experiments, the single-objective scheme stopped evolving at the $16$th generation on the MIR-1K dataset and  the $31$rd generation on the DSD100 dataset when the SDR of the best gene has no improvement for $S=8$ consecutive generations. For the  multi-objective scheme, we can observe the evolution process of all $100$ generations (i.e.,   $0$$\leq$G$\leq$$99$) on both MIR-1K  and DSD100 datasets. By comparing the two subfigures in  Fig.~\ref{mir1kgen} and   in Fig.~\ref{DSD100generations}, we can find that   the single-objective scheme and the multi-objective scheme  had  different  evolution trends.

As shown in Fig.~\ref{fig_mir1k_nsga2} and Fig.~\ref{fig_dsd2_nsga2},  the multi-objective scheme pushed the genes  toward  the Pareto front  generation by generation during the evolution process. In each generation,   a  set of  genes with different model complexities  and  SDR fitnesses were obtained.    More specifically, we can see that the  seed gene (represented by the black inverted triangle) had a relatively high model complexity (Params$=$$2.33$ M) and a low SDR score ($9.5$ dB for  MIR-1K and $5.9$ dB  for  DSD100). As the evolution proceeded, the new generations  gradually moved  toward the Pareto optimal front. For example, the first $10$ generations in  Fig.~\ref{fig_mir1k_nsga2} and Fig.~\ref{fig_dsd2_nsga2} (red and yellow points)  spread widely,   the generations from $10$ to $40$ (yellow and green points) started to move   to the lower-right boundary,  and  the higher generations, e.g., $70$ to $99$ generations (blue and pink points),  converged  to  the Pareto optimal front approximately. These results suggested that we could  obtain better genes (in model complexity, in SDR performance, or in both) as the evolution proceeded. Finally,  a set of structures with  better overall performance in model complexity and/or SDR performance were obtained, which can deal with different complexity requirements.

Compared with the  multi-objective scheme,  the model complexity of genes in the single-objective scheme was not reduced during the evolution process,  as shown in Fig.~\ref{fig_mir1k_single} and Fig.~\ref{fig_dsd2_single}. This is because  the model complexity was not considered in the single-objective scheme. In particular, we can see from Fig.~\ref{fig_mir1k_single} and Fig.~\ref{fig_dsd2_single} that the single-objective scheme, without the constraint of model complexity, could steadily improve the SDR performance generation by generation. While  in the multi-objective scheme, the genes of  one generation (Fig.~\ref{fig_mir1k_nsga2} and Fig.~\ref{fig_dsd2_nsga2}) showed much  difference in SDR performance (so that they can  deal with different complexity levels). In addition, by comparing Fig.~\ref{fig_mir1k_nsga2} with Fig.~\ref{fig_mir1k_single}, and Fig.~\ref{fig_dsd2_nsga2}  with Fig.~\ref{fig_dsd2_single}, we can   find that  the single-objective scheme could achieve a similar SDR performance to  the multi-objective scheme with much fewer generations. For example  in Fig.~\ref{fig_mir1k_single}, the single-objective scheme reached a SDR $=11$ dB using  only  $5$$\leq$G$\leq$$10$ generations, while this  required  at least  $20$ generations in the  multi-objective scheme. Nevertheless, we can  observe that the multi-objective scheme could achieve a  lower model complexity at SDR $=11$ dB compared with the  single-objective scheme.   It is also  found that the single-objective scheme behaved differently  on  two datasets. On  the MIR-1K dataset (see Fig.~\ref{fig_mir1k_single}),  the model complexity was significantly  improved at high SDR score while this phenomenon was not apparent on the DSD100 dataset (see Fig.~\ref{fig_dsd2_single}).

We also labelled  some  representative   genes in Fig. 3 and Fig. 4 (see the legend in each subfigure). For the  multi-objective scheme in   Fig.~\ref{fig_mir1k_nsga2} and Fig.~\ref{fig_dsd2_nsga2},   the seed gene,   genes of  early generations (G$=$$25$ and G$=$$50$), and  genes of  the final genertion (G$=$$99$) are plotted. It is clear that better genes (in model complexity, in SDR performance, or in both) can be obtained during the evolution process. For the single-objective scheme, we  intentionally continued the evolution process for a few more generations. Typical  genes including the seed gene, genes of early generations (G$=$$1$, G$=$$8$ for MIR-1K and  G$=$$1$, G$=$$8$,  G$=$$16$ for DSD100), genes of  the final generation (G$=$$16$ for MIR-1K and G$=$$31$ for DSD100),  and genes after the final generation (G$=$$29$ for MIR-1K and G$=$$49$ for DSD100) are plotted in  Fig.~\ref{fig_mir1k_single} and Fig.~\ref{fig_dsd2_single}.  It is found from  Fig.~\ref{fig_mir1k_single} that the gene in later generation, i.e., S-29-1-MIR,  provided  higher SDR performance  than  the  best gene obtained  in the  evolution   process, i.e.,  S-16-1-MIR, on  the testing subset $\mathscr{D}_{te}$. For  DSD100 in Fig.~\ref{fig_dsd2_single}, the gene after the  final generation, i.e., S-49-1-DSD,   provided similar SDR performance to the final evolved genes, i.e., S-31-1-DSD and S-31-2-DSD.  The performance of all these  evolved genes was evaluated and compared  using the full training and testing datasets, as shown in the next section.
  
  \vspace{-2mm}  
\subsection{Final evaluations}\label{expresults2}
In this section, we compare the evolved structures and other SOTA MSVS methods using the full MIR-1K and DSD100  datasets. In accordance with previous methods~\cite{le2015deep, jeong2017singing,nugraha2016multichannel,uhlich2017improving, takahashi2017multi},  on the DSD100 dataset, we computed SDRs/SIRs/SARs of all songs and then computed their median values. 

\subsubsection{Quantitative evaluations} 
The  evolved structures   in  Fig.~3 and Fig.~4  were first compared with some  typical MR-CNN based   MSVS methods on the T-F masking framework: MR-FCNN~\cite{DMP2}, SHN~\cite{SHN}, and SA-SHN~\cite{weitao1}. The performances of all structures were evaluated   with respect to computational efficiency and separation performance. The  results on computational efficiency  are listed in  Table~\ref{tab:modelcomplexity} and  the    results on separation performance  are listed in Table~\ref{MIR1Kdata} (for MIR-1K dataset) and Table~\ref{DSD100data} (for DSD100 dataset). In these Tables,  we use SHN-$n$ and SA-SHN-$n$  to represent the $n$-layer SHN and  $n$-layer SA-SHN,  respectively.

\begin{table}[t] 
\caption{Computational efficiency of the proposed method (Seed, M-$\ast$, and S-$\ast$) and the existing methods (MR-FCNN, SHN-$\ast$, and SA-SHN-$\ast$).} \label{tab:modelcomplexity}
\center
  \setlength
  {\tabcolsep}{0.7em}
\renewcommand\arraystretch{0.95}
\begin{tabular}{|l|c| c|c|c| }
\hline
\multirow{2}{*} {Method} & Params & FLOPs & Training Speed  & Inferring Speed \\  
 & [M] &   [G] & [bat./s] &   [bat./s] \\ 
\hline
Seed & 2.33 & 129.72 & 31.61 & 93.09\\
\hline
M-25-5-MIR & 1.21 & 39.33 & 57.19 & 194.54\\
M-50-2-MIR & 0.37 & 18.27 & 97.50 & 349.71\\
M-50-5-MIR & 1.04 & 52.30 & 53.13 & 171.80\\
M-99-2-MIR & 0.30 & 11.64 & 121.37 & 445.10\\
M-99-4-MIR & 1.24 & 48.69 & 50.76 & 171.02\\
M-99-5-MIR & 2.42 & 130.66 & 31.58 & 94.34\\
M-99-8-MIR & 8.27 & 454.28 & 13.13 & 35.78\\
\hline

M-25-4-DSD & 0.77 & 26.91 & 71.30 & 249.70\\
M-50-8-DSD & 2.73 & 139.41 & 29.21 & 87.59\\
M-99-2-DSD & {0.15} & {7.91} & {175.22} & {621.22}\\
M-99-4-DSD & 0.38 & 15.64 & 108.06 & 408.32\\
M-99-6-DSD & 0.59 & 21.63 & 87.76 & 311.76\\
M-99-7-DSD & 3.18 & 151.41 & 27.83 & 82.70\\

\hline
S-1-1-MIR & 2.47 & 135.31 & 30.40 & 89.67\\
S-8-1-MIR & 3.91 & 193.09 & 23.09 & 67.28\\
S-16-1-MIR & 6.76 & 404.45 & 13.89 & 38.72\\
S-29-1-MIR & 6.67 & 400.50 & 13.90 & 38.79\\
\hline
S-1-1-DSD & 2.55 & 135.85 & 30.16 & 89.48\\
S-8-1-DSD & 2.80 & 138.79 & 29.87 & 88.55\\
S-16-1-DSD & 2.53 & 136.47 & 30.45 & 91.05\\
S-31-1-DSD & 2.15 & 116.94 & 33.62 & 102.16\\
S-31-2-DSD & 2.84 & 144.63 & 29.13 & 85.66\\
S-49-1-DSD & 2.48 & 125.90 & 32.72 & 97.97\\
\hline

\hline
MR-FCNN & 0.56 & 36.56 & 9.03 & 18.59\\

SHN-1 & 9.06 & 168.29 & 29.94 & 87.70\\
SHN-2 & 17.46 & 292.87 & 16.70 & 49.19\\
SHN-4 & 34.18 & 537.66 & 8.84 & 26.09\\

SA-SHN-1 & 9.85 & 197.29 & 14.41 & 40.08\\
SA-SHN-2 & 19.03 & 350.87 & 7.56 & 20.95\\
SA-SHN-4 & 37.33 & 653.67 & 3.87 & 10.70\\

\hline
\end{tabular} 
\vspace{-2mm}

\end{table}

\textbf{Computational efficiency:} The  computational efficiency in Table~\ref{tab:modelcomplexity} was calculated  in  theory and measured in real hardware/software environment. The theoretical efficiency was given by Params  and FLOPs, where Params denotes the number of trainable parameters of  each structure and FLOPs represents the number of floating-point operations for testing (inferring)  in one batch.
In practice, two structures with similar Params and FLOPs may have  different computation speeds, thus  the computational efficiency was also  measured in  real  hardware/software environment\footnote{The GPU is RTX 2080Ti, CPU is Intel Core
i9 9900K, and the  memory is 4$\times$16G DDR4 (3200 MHz). In Linux operating system, we use  TensorFlow 2.0 with CUDA 10.1 and cuDNN 7.6.}. The real computational efficiency in training and  inferring was given  in bat./s. that is, the number of batches per second.

According to Table~\ref{tab:modelcomplexity},  the multi-objective scheme  provided multiple   structures with varying model complexities in one generation, e.g., M-99-2/4/5/8-MIR. In particular,  most evolved structures in the multi-objective scheme had a lower model complexity   than the seed on both datasets.  For single-objective scheme,  the model complexity of the evolved structures on the MIR-1K was increased generation by  generation and most structures had a slightly  higher model  complexity than the seed. On DSD100, however, the increase in the model complexity was not apparent during the evolution process.

\begin{table}[t] 
\vspace{-2mm}
\caption{The separation performance on MIR-1K (in dB)  of the proposed method (Seed, M-$\ast$, and S-$\ast$) and the existing SOTA methods (MR-FCNN, SHN-$\ast$, and SA-SHN-$\ast$).} \label{MIR1Kdata}
\vspace{-2mm}
\center
  \setlength
  {\tabcolsep}{0.3em}
\renewcommand\arraystretch{1}

\begin{tabular}{|l|c| c|c|c|c|c|c| c|c| }
\hline

\multirow{2}{*}{Method} & \multicolumn{3}{c|}{Acc} & \multicolumn{3}{c|}{Vocal} & \multicolumn{3}{c|}{Mean}\\
\cline{2-10}
& {\tiny \bf GNSDR} & {\tiny \bf GSIR} & {\tiny \bf GSAR} & {\tiny \bf GNSDR} & {\tiny \bf GSIR} & {\tiny \bf GSAR} & {\tiny \bf GNSDR} & {\tiny \bf GSIR} & {\tiny \bf GSAR} \\ \hline

Seed & 10.23 & 14.16 & 13.08       		& 11.26 & 17.29 & 12.94 & 10.74 & 15.72 & 13.01\\ \hline
M-25-5-MIR  & 10.03 & 13.24 & 13.56 	&  11.80 &  {\bf18.95}  & 13.11 & 10.92 & 16.10 & 13.33\\
M-50-2-MIR  & 10.20 & 14.00 & 13.25 	& 11.41 & 17.56 & 13.05 & 10.80 & 15.78 & 13.15\\
M-50-5-MIR  & 10.41  &  {\bf14.85}& 12.97 	& 11.42 & 17.88 & 12.94 & 10.91 & {\bf16.37}  & 12.96\\
M-99-2-MIR & 10.13 & 14.04 & 13.09 	& 11.26 & 17.34 & 12.92 & 10.69 & 15.69 & 13.00\\
M-99-4-MIR  & 10.04 & 13.67& 13.25 	& 11.42 & 17.80 & 12.99 & 10.73 & 15.74 & 13.12\\
M-99-5-MIR  & 10.25 & 13.94  & 13.38	 & 11.54 & 17.39 & 13.28 & 10.90 & 15.66 & 13.33\\ 
M-99-8-MIR  & 10.31 & 13.68 &  13.68 	& {\bf11.89}  & 17.89 & 13.55 &  11.10& 15.78 & 13.62 \\\hline

S-1-1-MIR & 9.84 & 12.71 & 13.71 		& 11.69 & 18.03 & 13.22 & 10.76 & 15.37 &  13.47 \\
S-8-1-MIR  & 10.16 & 13.27 &  {\bf13.77}  	&  11.85  & 18.02 &  13.45  & 11.00 & 15.64 &  13.61 \\
S-16-1-MIR  &  {\bf10.55}  & 14.18 & 13.65 	& {\bf11.89} & 17.80 &  {\bf13.60}  &  {\bf11.22} & 15.99 & {\bf13.63} \\
S-29-1-MIR  &  10.51  & 14.20 &  13.58 	& 11.83 & 17.79 &  13.51  &  11.17 & 16.00 & 13.54 \\ 

\hline

\hline
MR-FCNN & 8.65 & 11.65 & 12.35 		& 9.66 & 15.72 & 11.40 & 9.16 & 13.68 & 11.87\\
\hline
SHN-1 & 9.85 & 13.66 & 12.85 			& 10.88 & 16.63 & 12.71 & 10.36 & 15.15 & 12.78\\
SHN-2 & 9.94 & 13.67 & 12.96 			& 11.10 & 17.13 & 12.82 & 10.52 & 15.40 & 12.89\\
SHN-4 & 9.97 & 13.65 & 13.08 			& 11.13 & 17.09 & 12.89 & 10.55 & 15.37 & 12.98\\
\hline
SA-SHN-1 & 10.12 & 13.78 & 13.25 		& 11.32 & 17.15 & 13.10 & 10.72 & 15.47 & 13.18\\
SA-SHN-2 & 10.34 & 13.99 & 13.46 		& {11.71} & 17.58 & {13.44} & {11.02} & 15.79 & {13.45}\\
SA-SHN-4 & {10.53} & 14.54 & 13.38 	& {11.75} & 17.87 & 13.40 & {11.14} & {16.21} & 13.39\\

\hline
\end{tabular}   
\vspace{-2mm}
 \end{table}

The theoretical model complexity of  MR-FCNN was lower than those of the seed and some of the evolved structures (see Params and FLOPs). However,  in real  environment,  its computation speed was much  slower than the seed and  the evolved structures, e.g., MR-FCNN vs. S-8-1-MIR, MR-FCNN vs. M-50-8-DSD. In particular, we can also find that some evolved structures, e.g.,   M-50-2-MIR, M-99-2-MIR, and M-99-2-DSD, could  achieve  lower  theoretical model complexity than MR-FCNN. In SHN and  SA-SHN, the model complexity was increased with layer number and  the model complexities of these  two methods were much higher  than  those of the seed, the  multi-objective scheme,  the single-objective scheme, and the MR-FCNN.

\textbf{Separation performance:} 
We  can see from Table~\ref{MIR1Kdata} (MIR-1K dataset) that the evolved structures in both single-objective and multi-objective  schemes achieved  higher GNSDR and GSIR performance  on the Vocal source and higher  GSAR performance  on the Acc source than the seed. For DSD100 in Table~\ref{DSD100data}, most evolved structures achieved higher  SDR performance on Acc and Vocal sources than the seed. For Vocal source, most evolved structures achieved  higher  SIR/SAR performance. The last three columns of Table~\ref{MIR1Kdata} and Table~\ref{DSD100data} listed the mean   results of Vocal and Acc. One can see that the overall separation performances of most evolved structures in  single-objective and multi-objective  schemes outperform  the seed in three evaluation metrics. In addition, by comparing the proposed method (including the seed, the single-objective scheme,  and the multi-objective scheme) with other methods, one can see that the single-objective scheme,   the multi-objective scheme, and the SA-SHN outperformed the MR-FCNN and the  SHN.

\textbf{Computational efficiency vs. Separation performance:} 

\textit{(i) Proposed method:} By comparing Table~\ref{tab:modelcomplexity} and the mean   results in Tables~\ref{MIR1Kdata}-\ref{DSD100data}, we can find  that within the same generation of the multi-objective scheme, the structures with a higher model complexity can provide higher  performance on both datasets, e.g. from M-99-2-MIR to M-99-8-MIR,  from M-99-2-DSD to M-99-7-DSD. In  single-objective scheme, a higher generation (with increased model complexity) usually achieved  better  separation performance,  e.g. from S-1-1-MIR to S-16-1-MIR,   from S-1-1-DSD to S-31-1-DSD.  In particular,   according to Fig.~\ref{fig_mir1k_single}, the  S-29-1-MIR (a structure of  later generation  after the stopping criterion was satisfied) provided  higher SDR performance  than  the  final evolved gene  S-16-1-MIR on  the testing subset $\mathscr{D}_{te}$, while according to Tables~\ref{MIR1Kdata}, this gene does not outperform the S-16-1-MIR  on the full MIR-1K dataset. This result verified the effectiveness of our stopping criteria of the single-objective scheme.

\begin{table}[t] 
\vspace{-2mm}
\caption{The separation performance on  DSD100  (in dB) of the proposed method (Seed, M-$\ast$, and S-$\ast$) and the existing SOTA methods (MR-FCNN, SHN-$\ast$, and SA-SHN-$\ast$).} \label{DSD100data}
\vspace{-2mm}
\center
  \setlength
  {\tabcolsep}{0.3em}
\renewcommand\arraystretch{1}
\begin{tabular}{|l|c| c|c|c|c|c|c| c|c| }
\hline
\multirow{2}{*}{Method} & \multicolumn{3}{c|}{Acc. (Median)} & \multicolumn{3}{c|}{Vocal (Median)} & \multicolumn{3}{c|}{Mean }\\
\cline{2-10}
& { SDR} & SIR & SAR & SDR & SIR & SAR & SDR & SIR & SAR \\ \hline

Seed & 12.18 & 18.36 & 14.47 & 5.47 & 13.16 & 7.01 & 8.83 & 15.76 & 10.74\\ \hline
M-25-4-DSD &{\bf 12.78} & 17.91 & 14.80 & 6.21 & 14.32 & 7.24 &  9.50  & 16.12 & 11.02 \\
M-50-8-DSD & 12.70 &  18.34  & 14.88  & 6.31 & 14.85 & 7.45 & 9.51 & 16.60 &11.16 \\
M-99-2-DSD & 11.96 & 18.40 & 13.95 & 5.36 & 13.10 & 6.53 & 8.66 & 15.75 & 10.24\\
M-99-4-DSD & 12.52 & 18.25 & 14.43 & 5.95 & 14.27 & 7.12 & 9.23 & 16.26 & 10.78\\
M-99-6-DSD &  12.64 & 18.09 & 14.70 & 6.15 & 14.53 & 7.25 &  9.39 & 16.31 & 10.98\\
M-99-7-DSD &  12.64  & 18.33 & 14.83 & {\bf 6.42}  & 14.79 & {\bf 7.51}  &  {\bf 9.53} & 16.56 &  {\bf 11.17} \\ \hline

S-1-1-DSD & 12.33 & {\bf 18.49} & 14.45 & 5.68 & 13.16 & 7.22 & 9.01 & 15.82 & 10.84\\
S-8-1-DSD & 12.39 & 17.78 & 14.44 & 5.82 & 14.56 & 7.04 & 9.11 & 16.17 & 10.74\\
S-16-1-DSD & 12.41 & 18.05 &  14.74  & 6.26 &  {\bf 15.24}  & 7.14 & 9.34 &{\bf 16.64}  & 10.94\\
S-31-1-DSD & 12.60 & 18.48 & 14.72 & 6.15 & 14.76 &  7.36 &  9.38 &  16.62  & 11.04\\
S-31-2-DSD&  12.70  & 18.28 & 14.69 & 6.24 & 14.68 & 7.31 &  9.47  & 16.48 & 11.00\\
S-49-1-DSD &  12.62 & 18.25 & 14.54 & 6.23 & 14.89 & 7.24 &  9.42 &  16.57  & 10.89\\ 
\hline

\hline

MR-FCNN & 11.28 & 16.48 & 13.59 & 4.76 & 12.43 & 5.83 & 8.02 & 14.45 & 9.71\\
\hline
SHN-1 & 12.11 & 17.78 & 14.20 & 5.42 & 13.46 & 6.66 & 8.76 & 15.62 & 10.43\\
SHN-2 & 12.01 & 17.95 & 14.43 & 5.67 & 13.80 & 6.76 & 8.84 & 15.88 & 10.60\\
SHN-4 & 12.17 & 17.63 & 14.61 & 5.85 & 14.29 & 7.07 & 9.01 & 15.96 & 10.84\\
\hline
SA-SHN-1 & 12.17 & 17.71 & {14.73} & 5.91 & 14.76 & 7.17 & 9.04 & 16.23 & 10.95\\
SA-SHN-2 & 12.33 & 18.06 & {14.73} & 6.11 & 14.79 & 7.27 & 9.22 & 16.42 & {11.00}\\
SA-SHN-4 & {12.63} & 18.04 & {\bf 14.90} & {6.24} & {15.14} & {7.31} & {9.43} & {16.59} & {11.10}\\
 
\hline
\end{tabular}   
\vspace{-2mm}
\end{table}

 \begin{figure*}
 \vspace{-8mm} 
\center
\subfigure[On MIR-1K dataset]{
\includegraphics[width=0.48\textwidth]{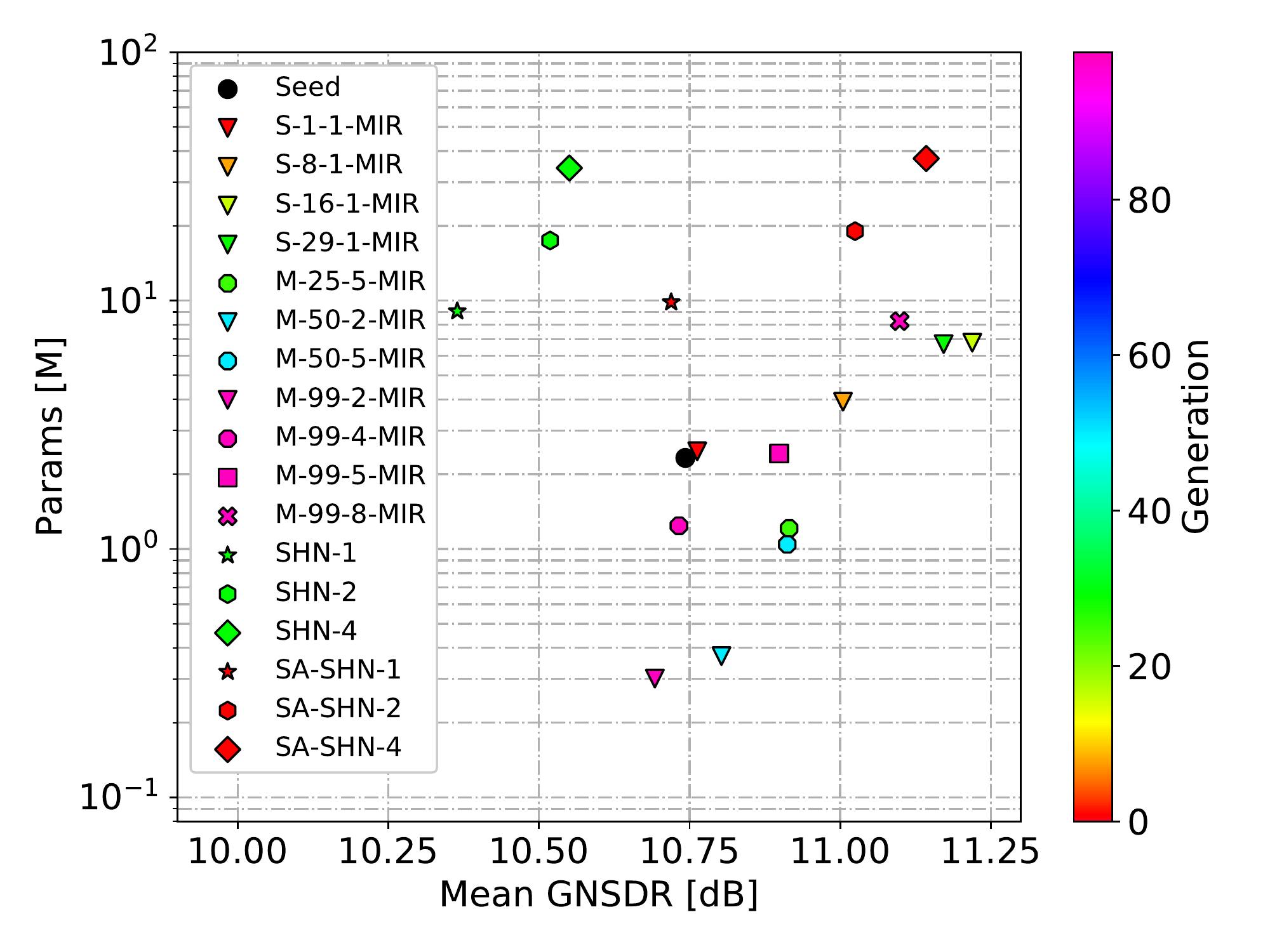} 
\label{fig_comp_all_dsd100}}
\subfigure[On DSD100 dataset]{
\includegraphics[width=0.48\textwidth]{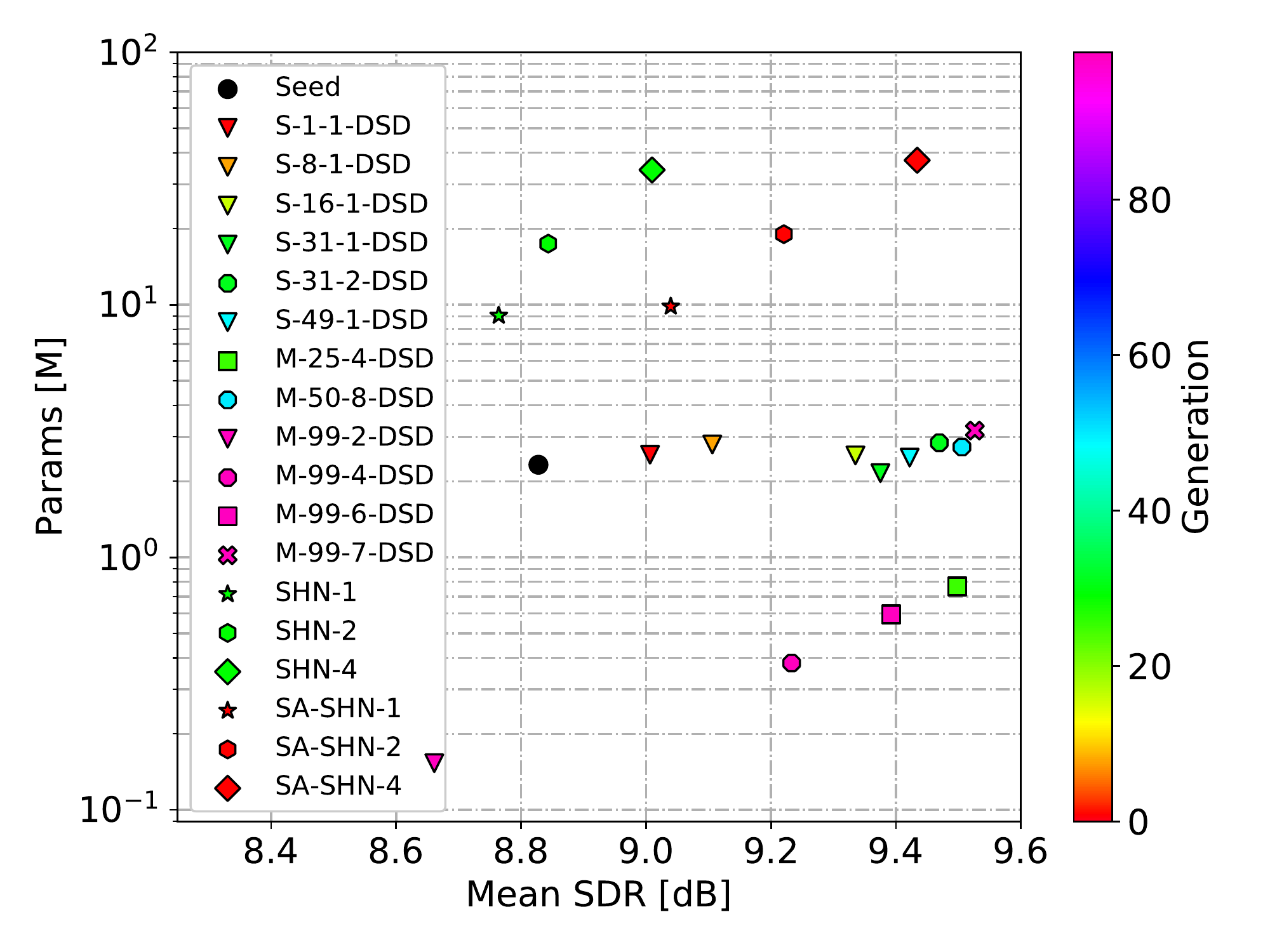} 
\label{fig_comp_all_mir1k}}
\vspace{-2mm}
\caption{Visualization of all structures in Params (vertical axis) and mean GNSDR/SDR (horizontal axis).}
\label{visualmodels}
\vspace{-2mm}
\end{figure*}

 \begin{figure*}[t]
\center
\subfigure[Vocal]{
 \includegraphics[width=0.48\textwidth]{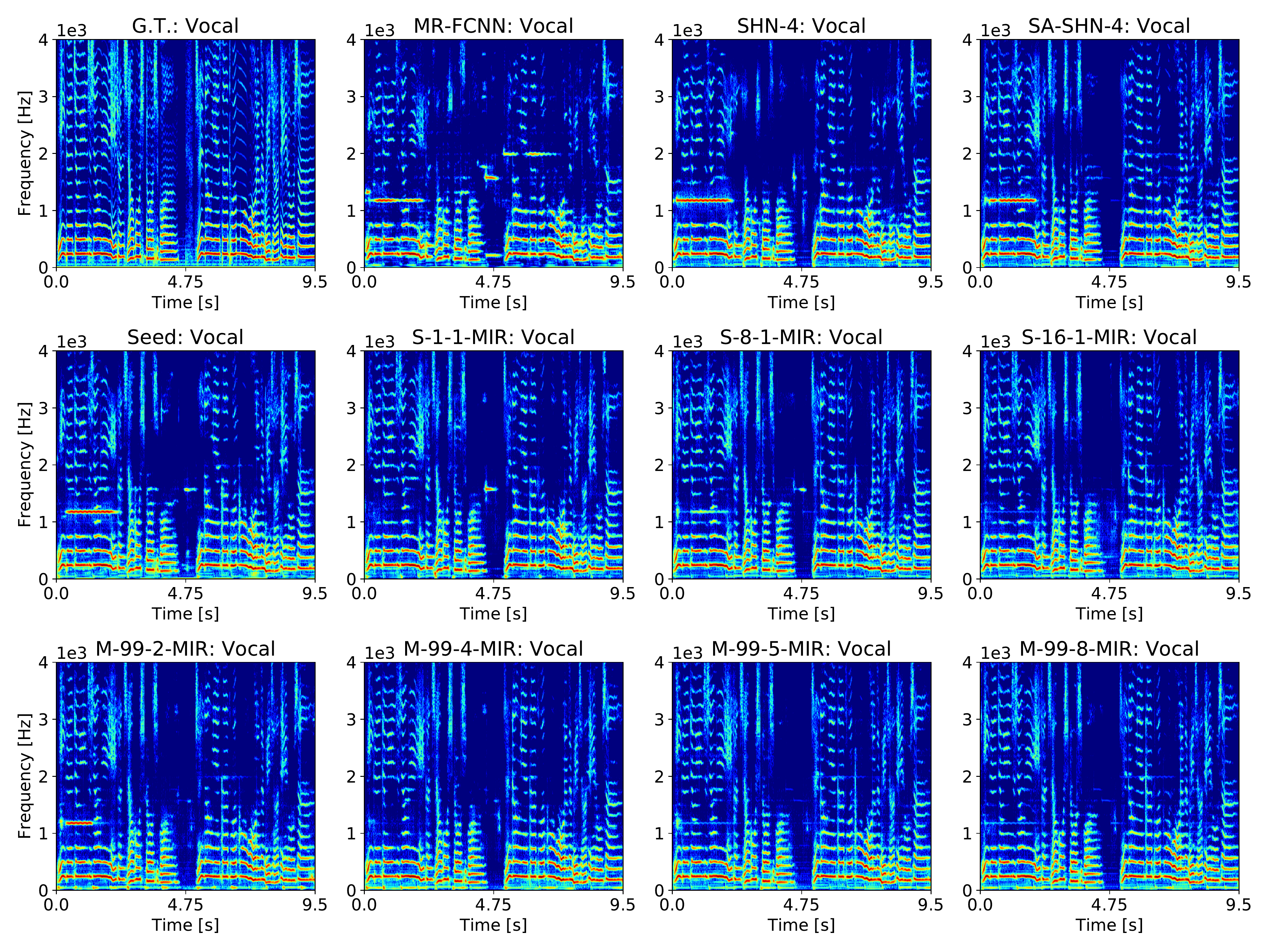}}
 \subfigure[Acc]{
 \includegraphics[width=0.48\textwidth]{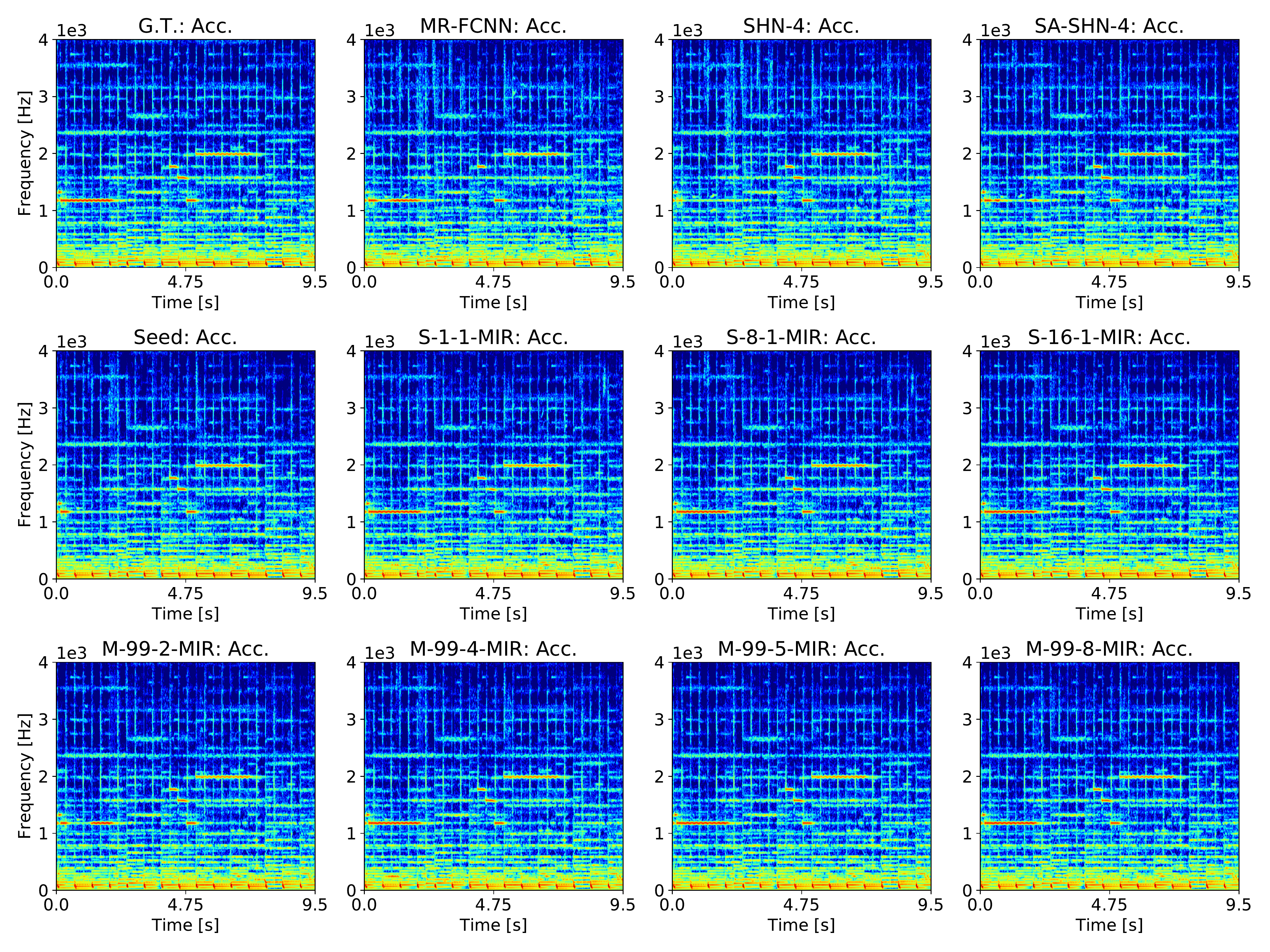}}
 \vspace{-2mm} 
\caption{Qualitative comparison of the proposed method with  MR-FCNN, SHN-4, and SA-SHN-4 on MIR-1K dataset.}
\label{spectrograms}
\vspace{-4mm}
\end{figure*}

By comparing the multi-objective scheme and  the seed, one can see  that the multi-objective  scheme can obtain better  separation results than the seed with a lower model complexity. For example,  the M-99-6-DSD achieved $0.56$ dB improvement on mean SDR than the seed using only $25.3$\% Params and  $16.7$\% FLOPs of the seed. In real environment,  this  structure was also  $2.77$ times (Training) and  $3.35$ times (Inferring) faster than  the Seed. When comparing the single-objective scheme with the seed, we can find that the single-objective scheme achieved much better separation results  with a slightly higher model complexity  on the MIR-1K dataset, e.g., the  S-16-1-MIR, which had $0.48$ dB improvement on  mean GNSDR than that of the seed with additional cost of $4.43$ M Params and $274.73$ FLOPs. On the DSD100 dataset, the single-objective scheme achieved much better separation results with similar  or even lower model complexity to the seed,  e.g.,  the S-31-1-DSD,  which obtained $0.55$ dB improvement on mean SDR and had a lower model complexity (only $2.15$  in Params and $116.94$ in FLOPs) than the seed.   
  
When comparing the single-objective scheme with  the multi-objective scheme, we can find that the multi-objective scheme could sometimes find more effective and efficient structures (similar or lower  model complexity but better separation performance) than the single-objective scheme. For example, the M-99-6-DSD is $0.01$ dB higher in the mean SDR than the S-31-1-DSD but with only $27.4$\% Params and $18.5$\% FLOPs of the S-31-1-DSD. In the real environment, the M-99-6-DSD was $2.61$ times (Training) and $3.05$ times (Inferring) faster than the S-31-1-DSD. Such phenomenon can also be observed on the MIR-1K dataset, e.g., the M-50-5-MIR with only $26.6$\% Params and $27.1$\% FLOPs of the S-8-1-MIR was only $0.09$ dB lower  than the S-8-1-MIR in  mean GNSDR. In the real environment, the M-50-5-MIR was $2.30$ times (training) and $2.55$ times (Inferring) faster than the S-8-1-MIR. These observations suggested  that  the multi-objective  scheme can greatly reduce the model complexity while maintaining acceptable separation performance.

\textit{(ii)  Proposed method vs. other methods:}
Compared with the proposed method (the seed, the single-objective scheme,  and the multi-objective scheme), the MR-FCNN has  lower theoretical model complexity ($0.56$  in Params and $36.56$ in FLOPs), while in real environment, it was much slower in training ($9.03$ bat./s) and inferring ($18.59$ bat./s) than the proposed method. For separation performance, the MR-FCNN was much worse than the proposed method. In particular,  we can see from Tables~\ref{MIR1Kdata}-\ref{DSD100data} that the evolved structures  in multi-objective scheme,  e.g., M-99-2-MIR, M-50-2-MIR, and M-99-2-DSD, could   achieve better separation performance in mean GNSDR/SDR than MR-FCNN even with  lower model complexity.

The  SHN and SA-SHN also achieved good separation performance, especially the SA-SHN. However, these two methods had low computational efficiency. For example, on the MIR-1K dataset, the SHN-4 (the best performance of SHN) and the M-99-2-MIR of the multi-objective scheme, have similar mean GNSDR results, while  the model complexity of SHN-4 was $113.9$ times (Params) and $46.2$ times (FLOPs)  of those of M-99-2-MIR. In real environment, the SHN-4 was $13.7$ times  (training) and $17.1$ times (Inferring) slower than  the M-99-2-MIR. Similar phenomenon can be also observed from, e.g.  SHN-4 vs. S-1-1-MIR, SHN-4 vs. S-1-1-DSD,  SHN-4 vs. M-99-6-DSD. For SA-SHN, we can see from  Tables~\ref{MIR1Kdata}-\ref{DSD100data} that when  the SA-SHN-4 (the best performance of SA-SHN) had similar GNSDR/SDR results  to  the  single-objective and  multi-objective schemes on the MIR-1K and DSD-1K datasets, its model 
complexity was much higher than the proposed structures, e.g., SA-SHN-4 vs. S-16-1-MIR, SA-SHN-4 vs. M-25-4-DSD.

All the above results suggested that the proposed method (especially the multi-objective scheme) was more effective and efficient than the MR-FCNN, the SHN, and the SA-SHN. In order to clearly visualize these quantitive results,  we plotted all the data (except for MR-FCNN) of Tables~\ref{tab:modelcomplexity}-\ref{DSD100data} in Fig.~\ref{visualmodels}, 
where the vertical axis is the Params and the horizontal axis is the mean GNSDR/SDR.

   \begin{table}[t]
    \vspace{-2mm}    
\caption{Comparison of the proposed method (Seed, M-$\ast$, and S-$\ast$) with other MSVS methods (MLRR, DRNN, ModGD, and U-Net) on MIR-1K dataset (in dB), where ``--'' means corresponding results were not provided by the method.}
 \begin{center}
 \center
  \setlength
  {\tabcolsep}{0.63em}
 \renewcommand\arraystretch{1} 
 
 \begin{tabular}{|c|c|c|c|c|c|c|}
  \hline
  \multirow{2}{*}{Method} & \multicolumn{3}{c|}{Vocal}&\multicolumn{3}{c|}{Acc}  \\
  \cline{2-7}
 & GNSDR & GSIR & GSAR& GNSDR & GSIR & GSAR\\
  \hline
MLRR~\cite{yang2013low}  & 3.85 & 5.63 & 10.70 & 4.19 & 7.80 & 8.22\\
DRNN~\cite{huang2015joint}  & 7.45 & 13.08 & 9.68 &--&--&--\\
ModGD~\cite{sebastian2016group} & 7.50 & 13.73 & 9.45&--&--&-- \\
U-Net~\cite{jansson2017singing} & 7.43 & 11.79 & 10.42& 7.45 & 11.43 & 10.41  \\ 
\hline

\hline
Seed   & 11.26 & 17.29 & 12.94 & 10.23            & 14.16             & 13.08 \\ \hline
M-25-5-MIR  &  11.80 &  {\bf 18.95}  & 13.11& 10.03   & 13.24   & 13.56\\
M-50-2-MIR & 11.41 & 17.56 & 13.05  &10.20   & 14.00   & 13.25 \\
M-50-5-MIR & 11.42 & 17.88 & 12.94 &  10.41 & {\bf 14.85}  & 12.97 \\
M-99-2-MIR & 11.26 & 17.34 & 12.92 & 10.13              & 14.04           & 13.09\\
M-99-4-MIR & 11.42 & 17.80 & 12.99 & 10.04             & 13.67            & 13.25\\
M-99-5-MIR  & 11.54 & 17.39 & 13.28 & 10.25   & 13.94   & 13.38 \\ 
M-99-8-MIR & {\bf 11.89} & 17.89 &  13.55   & 10.31   & 13.68   &13.68 \\ \hline

S-1-1-MIR  & 11.69 & 18.03 & 13.22& 9.84              & 12.71               &  13.71\\
S-8-1-MIR &  11.85 & 18.02 & 13.45 & 10.16   & 13.27   & {\bf 13.77}\\
S-16-1-MIR & {\bf 11.89} & 17.80 &{\bf 13.60} & {\bf 10.55}   & 14.18   &  13.65 \\
S-29-1-MIR &  11.83 & 17.79 &  13.51 &  10.51   & 14.20   &  13.58 \\       
\hline

\end{tabular}

\end{center}

 \label{Othercompare1}
 \vspace{-4mm} 
 
\end{table}

 \subsubsection{Qualitative Results}
We also qualitatively compared the separation performance of the above methods.  The separation results on an  exemplar MIR-1K song clip (\textit{geniusturtle\_6\_04}) are shown in Fig.~\ref{spectrograms}.  By comparing  the ground truth (G.T.) Vocal and Acc, one can see that the MR-FCNN, the SHN-4,  the  SA-SHN-4, and the seed of the proposed method wrongly assigned an important  frequency component of  Acc ($1200$ Hz appearing  around   $0$s$\mathtt{\sim}$$2$s)   to Vocal. Besides, the MR-FCNN and the SHN-4 could not capture some of the fine vocal details. In contrast, the evolved structures in single-objective scheme, e.g.,  S-1-1-MIR, S-8-1-MIR, and S-16-1-MIR,  correctly put  this frequency component back to the  Acc. In  multi-objective scheme,  the separation results of  several  structures in the $99$th  generation, e.g., M-99-2/4/5/8-MIR, are exhibited. It is shown that the  M-99-4/5/8-MIR correctly assigned the $1200$ Hz frequency component back  to Acc while the M-99-2-MIR did not.  According to   Table~\ref{tab:modelcomplexity},  we can find that the M-99-2-MIR compromised  the separation performance with  a very low model complexity. Finally, one can see that  the estimated magnitude spectrograms of the   Vocal and Acc obtained by M-99-4/5/8-MIR  were quite similar to those of the ground truth  Vocal and Acc sources.

\subsubsection{Comparsion with other  methods}
We finally compared the proposed method with   other MSVS methods. The results are listed in Tables~\ref{Othercompare1}-\ref{Othercompare2}. These numerical results verified the separation performance   of the proposed method.

 \begin{table}[t]
  \vspace{-2mm}  
 \caption{Median SDR values of the proposed method (Seed, M-$\ast$, and S-$\ast$) and other MSVS methods (DeepNMF, wRPCA, NUG, BLEND, and MM-DenseNet) on  DSD100 dataset (in dB).} 
 \begin{center}
  \setlength
  {\tabcolsep}{2.4em}

\renewcommand\arraystretch{0.9}
  \begin{tabular}{|c|c|c|}

  \hline
  Method & \;\;Vocal\;\; & Acc\\
  \hline
  DeepNMF~\cite{le2015deep} & 2.75 & 8.90 \\
  wRPCA~\cite{jeong2017singing} & 3.92 & 9.45 \\
  NUG~\cite{nugraha2016multichannel}  &  4.55  & 10.29 \\
  BLEND~\cite{uhlich2017improving}  &  5.23  & 11.70 \\
  MM-DenseNet~\cite{takahashi2017multi} & {6.00} & 12.10 \\
\hline

\hline
Seed  & 5.47   & 12.18      \\ \hline
M-25-4-DSD &6.21 &{\bf 12.78}\\
M-50-8-DSD& 6.31&12.70\\
M-99-2-DSD  & 5.36 & 11.96 \\
M-99-4-DSD  & 5.95   & 12.52          \\
M-99-6-DSD  & 6.15     &  12.64         \\
M-99-7-DSD  &  {\bf 6.42}   &  12.64          \\ \hline

S-1-1-DSD   & 5.68& 12.33\\
S-8-1-DSD & 5.82     & 12.39         \\
S-16-1-DSD  & 6.26    & 12.41      \\
S-31-1-DSD  & 6.15      & 12.60           \\
S-31-2-DSD & 6.24       &  12.70          \\
S-49-1-DSD & 6.23       &  12.62         \\
\hline
\end{tabular}
\end{center}
 \label{Othercompare2}
 \vspace{-4mm} 
 
\end{table}

 \vspace{-2mm} 
\section{Conclusions}\label{sec:concl}
As the first attempt in the field of MSVS, this paper proposed an evolutionary  framework, i.e., the E-MRP-CNN, to  automatically find effective  neural networks for MSVS.  The proposed E-MRP-CNN is based on a novel MR-CNN namely MRP-CNN, which utilizes various-size  average pooling operators  for feature extraction.  Compared with existing MR-CNNs, the MRP-CNN has a low computational complexity and can effectively extract multi-resolution features for MSVS. We derived the E-MRP-CNN using single-objective and multiple-objective genetic algorithms. The single-objective E-MRP-CNN considers only the separation performance while  the multi-objective E-MRP-CNN  considers both the separation performance and the model complexity, and thus it provides a set of solutions to handle different separation performance and/or   model complexity requirements. Experimental results on the MIR-1K and DSD100 datasets showed that the proposed method (especially the multi-objective scheme) is more effective and efficient than the SOTA MSVS methods, which verified the effectiveness of the proposed method.

 \vspace{-2mm} 
\section{Acknowledge} 
This work was supported by National Natural Science Foundation of China (No.~61902280, 61373104), Natural Science Foundation of Tianjin (No.~19JCYBJC15600). It was also supported by a Grant-in-Aid for Scientific Research (B) (No. 17H01761) and I-O DATA foundation.

\bibliographystyle{IEEEtran}
\bibliography{strings}

\clearpage


\end{document}